\documentclass[a4paper,11pt]{article}
\usepackage[utf8]{inputenc}

\usepackage{epsfig, subfig}
\usepackage[usenames]{color}
\usepackage[stable]{footmisc}
\usepackage{graphicx}
\usepackage{dsfont}
\usepackage{titlesec, setspace}
\usepackage{fullpage,graphicx}
\usepackage{enumerate}
\usepackage{blindtext}
\usepackage{abstract}

\usepackage{amsmath, amssymb, stmaryrd, mathrsfs, amsfonts, amssymb, amsbsy, bm, wasysym, setspace, cancel}
\usepackage{bbm, accents}
\usepackage[a4paper,top=2.5cm, bottom=2.6cm, left=2cm, right=2cm]{geometry}

\usepackage[square, numbers, comma, sort&compress]{natbib}

\def\tr{\operatorname{tr}}

\def\Div{\operatorname{Div}}
\def\Grad{\operatorname{Grad}}

\def \d {\text{ d}}

\DeclareMathSymbol{\widehatsym}{\mathord}{largesymbols}{"62}
\newcommand\lowerwidehatsym{%
  \text{\smash{\raisebox{-1.3ex}{%
    $\widehatsym$}}}}
\newcommand\fixwidehat[1]{%
  \mathchoice
    {\accentset{\displaystyle\lowerwidehatsym}{#1}}
    {\accentset{\textstyle\lowerwidehatsym}{#1}}
    {\accentset{\scriptstyle\lowerwidehatsym}{#1}}
    {\accentset{\scriptscriptstyle\lowerwidehatsym}{#1}}
}

\title{Biological Growth in Bodies with Incoherent Interfaces}
\author{Digendranath Swain and Anurag Gupta\thanks{ag@iitk.ac.in}}
\date{\small Department of Mechanical Engineering, Indian Institute of Technology Kanpur, 208016, India.}
\begin{document}

\maketitle

\begin{abstract}
A general theory of thermodynamically consistent biomechanical--biochemical growth in a body, considering mass addition in the bulk and at an incoherent interface, is developed. The incoherency arises due to incompatibility of growth and elastic distortion tensors at the interface. The incoherent interface therefore acts as an additional source of internal stress besides allowing for rich growth kinematics. All the biochemicals in the model are essentially represented by nutrient concentration fields, in the bulk and at the interface. A nutrient balance law is postulated which, combined with mechanical balances and kinetic laws, yields an initial-boundary-value problem coupling the evolution of bulk and interfacial growth, on one hand, and the evolution of growth and nutrient concentration on the other. The problem is solved, and discussed in detail, for two distinct examples: annual ring formation during tree growth and healing of cutaneous wounds in animals. 
\end{abstract}


\section{Introduction}
\label{sec:intro}
Biological growth necessarily involves mass addition in bodies leading to microstructural rearrangements and internal stress distributions \cite{taber95, jones12}. It can be classified as either volumetric, surface, or interfacial based on the nature of mass exchange with the external environment. Whereas mass is added in the bulk material during volumetric growth \cite{Goriely17, epstein2000, Rodriguez1994} (e.g., in soft tumorous and arterial tissues),  it accretes on to the free surface of the body during surface growth \cite{Skalak1997, dicarlo05, truskinovsky17} (e.g., in hard horn and bone tissues). On the other hand, mass addition can also happen at a material or a non-material interface within the body \cite{ciarletta12, wolff15, ganghoffer11}, as is the case with ring formation in trees, healing of cutaneous animal wounds, growth of animal nails, etc. In fact, interfacial growth models can also provide a viable framework for studying problems in surface growth, e.g., by considering the external source to be the bulk body on one side of the interface \cite{truskinovsky17} or by assuming the interface to be between a bulk substrate and a growing two-dimensional film \cite{kuhl13b,kuhl13a}. In this paper, we develop a general three-dimensional finite deformation thermodynamically consistent theory of biomechanical--biochemical growth in a body where mass is being added both in the bulk as well at an incoherent interface. The general theory is discussed in detail for two problems: ring formation during tree growth and cutaneous wound healing in animals. Whereas the former is dealt with assuming a linearized strain kinematics, the latter is solved using a finite deformation framework.  The considerations of incoherency at the interface and of coupled biomechanical--biochemical bulk--interfacial growth are the main novelties of our work. 
 
An interface is called \textit{incoherent} whenever the jumps in elastic (and growth) distortions across it are incompatible, i.e. not restricted to be of a rank-one form \cite{Gupta12}. Such jumps become sources of residual stress, in addition to those arising in the bulk of the body. They also lead to richer growth kinematics, since the body on one side of the interface can grow without any resistence from the other side. Such a situation is commonly seen in the shrink fit problems of solid mechanics \cite{moulton11b}. While a general theory of growth in bodies with incoherent interfaces is lacking in the literature, several specific applications have appeared recently. These include the role of incoherent skin-wound interface in the wound healing problems leading to instability of wound shape \cite{WuAmar15}, skin wrinkling \cite{swain15}, and cavitating wound \cite{swain16}. An incoherent interface also led to circumferential buckling in growing bilayer cylindrical tubes \cite{moulton11b}. The interface between a growing thin film over a growing substrate, as considered recently by Kuhl and coauthors \cite{kuhl13b,kuhl13a}, is also incoherent. 

The second aspect of our theory is to extend the work by Ambrosi and Guillou \cite{ambrosi07} (see also \cite{buskohl14,oller10}) to include biochemistry in growing bodies with interfaces. Towards this end, we postulate a global nutrient balance law and derive local equations for the evolution of nutrient concentration fields, both in the bulk and at the interface, driven by the nutrient flux as well as by the growth kinetics. Reciprocally, the growth evolution is affected by the concentration evolution and the elasticity of the body. Such a coupling between biomechanics and biochemistry is essential for a realistic modeling of biological growth processes. Another coupling incorporated in our model is that between bulk and interfacial growth. The latter provides boundary data for the bulk growth and, in turn, is affected by the bulk deformation and stress fields.
   
As examples of our theory, we first revisit the classical problem of tree growth due to annual ring formation \cite{archer86,fournier90}. We depart from the earlier works by considering a thermodynamically consistent interfacial growth framework and incorporating nutrient biochemistry. Moreover, unlike previous models, we include elasticity of the bark and a non-uniform ring size distribution in the trunk. Our approach provides a straightforward way to calculate the growth stress and nutrient concentration distributions in trees. Interestingly, we use bark elasticity to correlate the crack patterns on the bark with the growth strains therein. As the second example, we calculate the nutrient concentration field during the cutaneous wound healing process. This is done so as to achieve a better understanding of the nutrient chemistry in the problem, which can lead to efficient wound management and scar control. The biomechanical aspects of this problem were investigated recently by the present authors \cite{swain15, swain16}. 

The preliminaries for studying mechanics of incoherent interfaces are developed in Section \ref{sec:prelim}, following earlier work by one of the authors \cite{basak15a, basak15b, Gupta12}. In Section \ref{sec:balance_laws} we obtain the complete set of governing equations for the determination of deformation, stress, and nutrient concentration fields. These include the balance laws of mass, nutrient, and momentum, and the kinetic relations for interface migration, growth, and nutrient flux, both in the bulk and at the interface. The kinetic relations are consistent with the second law of thermodynamics. We also digress briefly to discuss growth of an elastic thin film over a growing elastic substrate. Analytically tractable models of tree growth and cutaneous wound healing are considered in Sections \ref{sec:rings} and \ref{wh}, respectively, and the proposed governing equations solved and discussed to illustrate the efficacy of our framework. We conclude our work in Section \ref{conc}.

\section{Preliminaries}
\label{sec:prelim}
Let $\mathcal{R}$ be the set of real numbers, $\mathcal{R}^+$ the set of positive real numbers, $\mathcal{V}$ the translation space (set of vectors) of a real three-dimensional Euclidean point space $\mathcal{E}$, and $Lin$ the set of second-order tensors consisting of all linear transformations from $\mathcal{V}$ to $\mathcal{V}$. The set of invertible, symmetric, symmetric positive-definite, and skew tensors are represented by $InvLin$, $Sym$, $Sym^+$, and $Skw$, respectively. The determinant, transpose, inverse, and cofactor of $\boldsymbol{A} \in Lin$ are denoted by $J_A$, $\boldsymbol{A}^{T}$, $\boldsymbol{A}^{-1}$, and $\boldsymbol{A}^*$, respectively. The identity tensor in $Lin$ is represented by $\boldsymbol{1}$. The Euclidean inner-product and the Euclidean norm in $Lin$ are defined as $\boldsymbol{A} \cdot \boldsymbol{B}=\tr(\boldsymbol{A}\boldsymbol{B}^T)$ and $|\boldsymbol{A}|^2=\boldsymbol{A} \cdot \boldsymbol{A}$, respectively, where $\boldsymbol{B} \in Lin$ and $\tr(\cdot)$ is the trace operator. We express the symmetric and skew-symmetric part of $\boldsymbol{A}$ as $\text{sym}(\boldsymbol{A})$ and $\text{skw}(\boldsymbol{A})$. The derivative of a continuously differentiable scalar-valued function of tensors $G(\boldsymbol{A})$ is denoted as $\partial_{\boldsymbol{A}} G \in Lin$, defined by
$G(\boldsymbol{A}+\boldsymbol{B})=G(\boldsymbol{A})+ \partial_{\boldsymbol{A}} G \cdot \boldsymbol{B} + o(|\boldsymbol{B}|)$,
where $o(|\boldsymbol{B}|)/{|\boldsymbol{B}|}\rightarrow 0$ when $|\boldsymbol{B}|\rightarrow 0$. Similar definitions hold for vector and tensor valued differentiable functions of scalars, vectors, and tensors.

\subsection{Deformation kinematics}
Let $\mathcal{B}_t \subset \mathcal{E}$ denote the current configuration of a growing body and let $\mathcal{B}_0 \subset \mathcal{E}$ be an arbitrary reference configuration such that there exists a bijective map $\boldsymbol{\chi}$ between $\mathcal{B}_0$ and $\mathcal{B}_t $. Assume $\mathcal{B}_t$ to be simply-connected. The position vector $\boldsymbol{x} \in \mathcal{B}_t$ is uniquely defined in terms of a position vector in the reference configuration $\boldsymbol{X} \in \mathcal{B}_0$, and time $t \in \mathcal{R}$, as $\boldsymbol{x}=\boldsymbol{\chi} (\boldsymbol{X}, t)$. The mapping $\boldsymbol{\chi}$ is assumed to be continuous but piecewise differentiable over $\mathcal{B}_0 $ and continuously differentiable with respect to $t$. The particle velocity and the deformation gradient are given by $\boldsymbol{v}=\dot{\boldsymbol{\chi}} \in \mathcal{V}$ and $\boldsymbol{F}=\Grad\boldsymbol{\chi}  \in InvLin$, respectively, where the superposed dot represents the material time derivative and $\Grad$ the gradient operator with respect to $\boldsymbol{X}$. The latter definition holds whenever $\boldsymbol{\chi}$ is differentiable at $\boldsymbol{X}$. Both $\boldsymbol{F}$ and $\boldsymbol{v}$ are assumed to be piecewise continuously differentiable over $\mathcal{B}_0$.

We consider a singular surface in the interior of $\mathcal{B}_0$, $\mathcal{I}_0= \{\boldsymbol{X} \in \mathcal{B}_0; \phi(\boldsymbol{X},t)=0\}$, where $\phi \in \mathcal{R}$ is a continuously differentiable level set function, see Figure \ref{decomp}. The unit normal $\mathbb{N}$ and the normal velocity $U$, associated with $\mathcal{I}_0$, are defined as $\mathbb{N}={\Grad \phi}/{|\Grad \phi|}$, and $U=-{\dot{\phi}}/{|\Grad \phi|}$, respectively. Various bulk fields, such as deformation gradient and stress, are allowed to be discontinuous in $\mathcal{B}_0$ only across $\mathcal{I}_0$. They are otherwise assumed to be smooth in $\mathcal{B}_0 / \mathcal{I}_0$. The projection tensor $\mathbbm{1}=\boldsymbol{1}- \mathbb{N} \otimes \mathbb{N} \in Sym$ project vectors onto the tangent space of the singular surface $\mathcal{I}_0$.
The jump and average of a piecewise continuous bulk field $\psi \in \mathcal{R}$ across $\mathcal{I}_0$ are given by $\llbracket \psi \rrbracket = \psi^+-\psi^-$ and $\langle \psi \rangle=(\psi^+ + \psi^-)/2$, respectively, where $\psi^+$ is the limiting value of $\psi$ as it approaches $\mathcal{I}_0$ from the bulk side into which $\mathbb{N}$ points and $\psi^-$ is the limiting value when approached from the other side of the interface. 
The interfacial fields $\mathbbm{g} \in \mathcal{R}$, $\mathbbm{v} \in \mathcal{V }$, and $\mathbb{G} \in Lin$, defined on $\mathcal{I}_0$, are differentiable at $\boldsymbol{X} \in \mathcal{I}_0$ if they have extensions ${g}  \in \mathcal{R}$, $\boldsymbol{v}  \in \mathcal{V}$, and $\boldsymbol{G} \in Lin$, to a neighborhood $\boldsymbol{X} \in \mathcal{B}_0$, which are differentiable at $\boldsymbol{X}$. The surface gradients of $\mathbbm{g}$, $\mathbbm{v}$, and $\mathbb{G}$ are then defined by $\Grad^S \mathbbm{g} = \mathbbm{1}(\Grad g)$, $\Grad^S \mathbbm{v} = (\Grad \boldsymbol{v}) \mathbbm{1}$, and $\Grad^S \mathbb{G} = (\Grad \boldsymbol{G}) \mathbbm{1}$.
The corresponding surface divergences are $\Div^S \mathbbm{v} = \tr(\Grad^S \mathbbm{v})$ and $\boldsymbol{k} \cdot \Div^S \mathbb{G} = \Div^S(\mathbb{G}^T \boldsymbol{k})$,
where $\boldsymbol{k} \in \mathcal{V}$ is fixed. The surface Laplacian of $\mathbbm{g}$ is given by $\Delta^S \mathbbm{g}=\Div^S(\Grad^S \mathbbm{g})$. The curvature tensor $\mathbb{L} \in Sym$ and the mean curvature ${\kappa} \in \mathcal{R}$
associated with $\mathcal{I}_0$ are defined as $\mathbb{L}=-\Grad^S \mathbb{N}$ and $\kappa=\tr \mathbb{L}$, respectively.
The normal time derivative of the interfacial field $\mathbbm g$,  continuously differentiable over $\mathcal{I}_0$, represents the rate of change of $\mathbbm g$ as observed by an observer sitting on the moving interface $\mathcal{I}_0$. It is defined in terms of its extension $g$ as
\begin{equation}\label{eq:prel10}
\mathring{\mathbbm g}=\dot{g}+U (\Grad g) \cdot \mathbb{N}.
\end{equation}
Using this definition, we can immediately deduce $\mathring{\mathbb{N}}=-\Grad^S U$.
The surface deformation gradient and the normal material velocity associated with $\mathcal{I}_0$, such that $\llbracket \boldsymbol{\chi} \rrbracket=\boldsymbol{0}$ for all $\boldsymbol{X} \in \mathcal{I}_0$, are given by \cite{Gupta12}
\begin{equation}\label{eq:prel14}
 \mathbb{F}=\Grad^S \boldsymbol{\chi}= \boldsymbol{F}^{\pm} \mathbbm{1} ~ \text{and}~ \mathbbm{v}=\mathring{\boldsymbol{\chi}}=\langle \boldsymbol{v} \rangle +U \langle \boldsymbol{F} \rangle  \mathbb{N},
\end{equation}
respectively. Clearly, $J_{\mathbb{F}}=0$, ${\mathbb{F}} {\mathbb{N}}=\boldsymbol{0}$, and ${\mathbb{F}} {\mathbbm{1}}=\mathbb{F}$. Also, as is well known, $\llbracket \boldsymbol{F} \rrbracket \mathbbm{1}=\boldsymbol{0}$ and $\llbracket \boldsymbol{v} \rrbracket + U \llbracket \boldsymbol{F} \rrbracket  \mathbb{N}=\boldsymbol{0}$. Velocity $\mathbbm{v}$ is the intrinsic material velocity of the particle points which coincide with the interface at time $t$. The surface gradient of $\mathbbm{v}$ is related to the normal time derivative of ${\mathbb{F}}$ as $\Grad^S \mathbbm{v} = \mathring{\mathbb{F}}  \mathbbm{1} - U {\mathbb{F}} \mathbb{L}$ \cite{Gupta12}. The ratio of infinitesimal surface areas (over the singular surface) in the current and the reference configuration is given by $j =|\mathbb{F}^* \mathbb{N}|$.

\subsection{Growth kinematics}
\begin{figure}[t!]
 \centering \includegraphics[width=0.50\textwidth, angle=270]{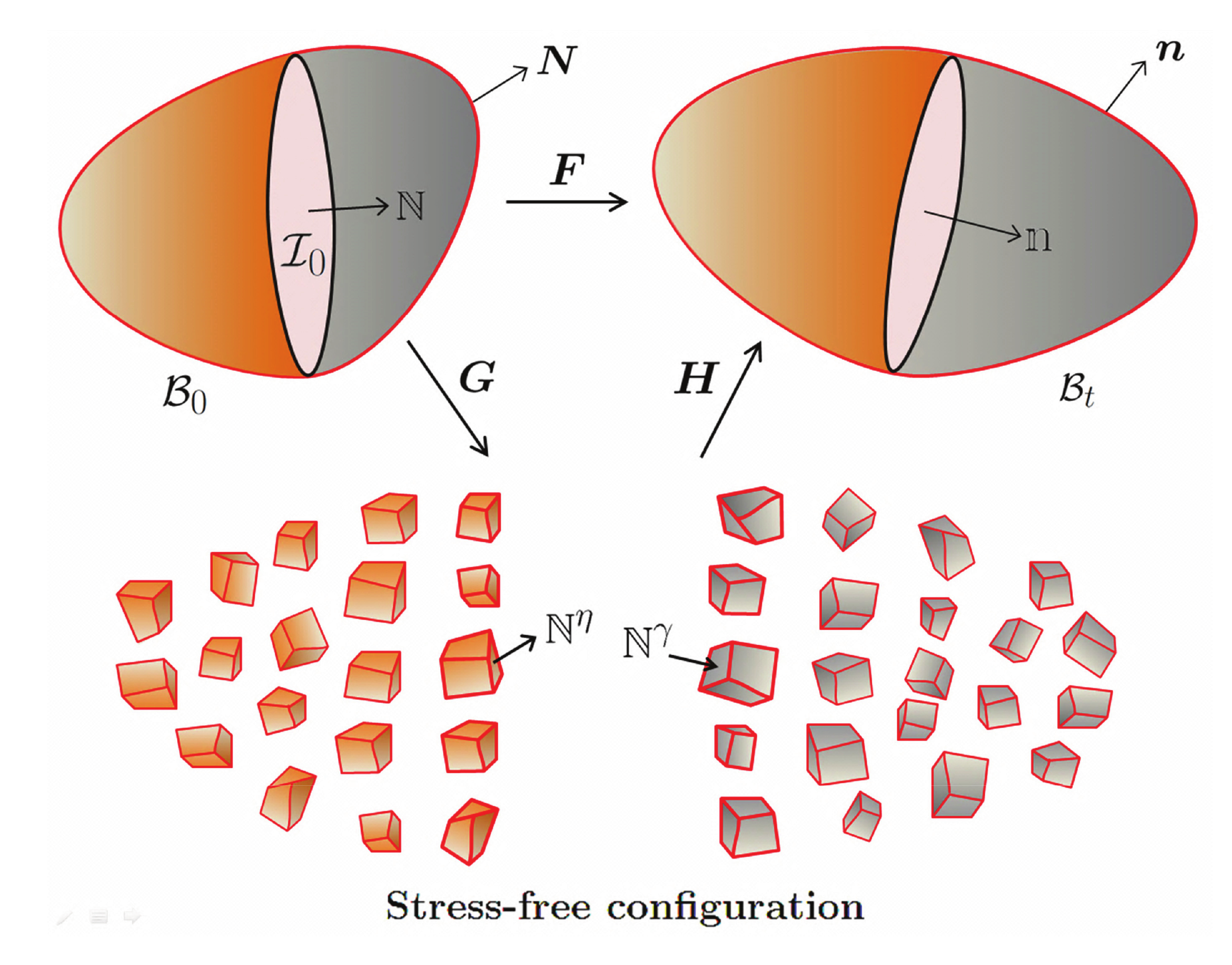}
  \caption{The reference, current, and stress-free configurations. The latter is obtained from the current configuration via elastic relaxation. The singular interface is incoherent yielding distinct normals $\mathbb{N}^\gamma$ and $\mathbb{N}^\eta$ in the relaxed configuration, as mapped from the same normal $\mathbb{N}$ (or $\mathbbm{n}$) in the reference (or current) configuration.}
  \label{decomp}
\end{figure}
Central to our biomechanical theory of growth is the multiplicative decomposition of the deformation gradient \cite{Rodriguez1994},
\begin{equation}
\boldsymbol{F}=\boldsymbol{HG} ~ \text{in}~ \mathcal{B}_0/\mathcal{I}_0, \label{dai2}
\end{equation}
where ${\boldsymbol H} \in InvLin$ is the elastic distortion tensor and ${\boldsymbol G} \in InvLin$ is the growth tensor, see Figure \ref{decomp}. The elastic distortion ${\boldsymbol H}$ represents in effect an elastic unloading of the body in the grown configuration $\mathcal{B}_t$ to a stress-free configuration assuming that the stress is purely elastic in origin. The stress-free configuration will not evolve unless the body grows. The tensor ${\boldsymbol G}$, which connects the stress-free configuration to the fixed reference configuration $\mathcal{B}_0$, hence represents the state of growth. The nature of the stress-free configuration, and hence of the elastic and growth distortion tensors, is governed by the choice of elastic response that is prescribed for the body. For instance, it is unique, modulo rigid body transformations, only for convex elastic energies \cite{Guptaetal07}. Due to its construction, the stress-free configuration is, in general, a disjoint set of disconnected domains in the Euclidean space. It can however be interpreted as a connected set in a non-Euclidean space which admits a non-metric affine connection \cite{Ayan16}. 

The multiplicative decomposition on the interface can be obtained by projecting  the limiting values of \eqref{dai2}, as the interface is approached, onto the interface $\mathcal{I}_0$.  We define the surface distortion tensors as $\mathbb{H}^\gamma = {\boldsymbol H}^+ \mathbbm{1}^\gamma, ~\mathbb{H}^\eta = {\boldsymbol H}^- \mathbbm{1}^\eta, ~\mathbb{G}^\gamma = {\boldsymbol G}^+ \mathbbm{1}, ~\text{and} ~\mathbb{G}^\eta = {\boldsymbol G}^- \mathbbm{1}$,
where the superscripts $\gamma$ and $\eta$ denote the two distinct surfaces in the stress-free configuration both related to the single interface in $\mathcal{B}_0$ or $\mathcal{B}_t$. The relaxation of the interface into two distinct surfaces is a consequence of the incoherency of the interface \cite{Gupta12, basak15a}. For a coherent interface, $\mathbb{G}^\gamma = \mathbb{G}^\eta$, or equivalently $\mathbb{H}^\gamma = \mathbb{H}^\eta$; the jumps in $ {\boldsymbol G}$ and ${\boldsymbol H}$ are then necessarily rank-one. In the preceding definitions we have used the projection tensors  $\mathbbm{1}^\gamma = {\boldsymbol 1} - {\mathbb N}^\gamma \otimes {\mathbb N}^\gamma~\text{and}~\mathbbm{1}^\eta = {\boldsymbol 1} - {\mathbb N}^\eta \otimes {\mathbb N}^\eta$,
where ${\mathbb N}^\gamma \in \mathcal{V}$ and ${\mathbb N}^\eta \in \mathcal{V}$ are unit normals associated with the two surfaces in the relaxed configuration such that
\begin{equation}
{\mathbb N}^\alpha = \frac{({\boldsymbol G}^\pm)^{-T} {\mathbb N}}{|({\boldsymbol G}^\pm)^{-T} {\mathbb N}|}, \label{epd5}
\end{equation}
with superscript $+$ appearing with $\alpha = \gamma$ and $-$ with $\alpha = \eta$. The normals ${\mathbb N}^\gamma$ and ${\mathbb N}^\eta$ coincide for coherent interfaces. The multiplicative decomposition on the incoherent interface, therefore, is of the form
\begin{equation}
\mathbb{F}  = \mathbb{H}^\alpha \mathbb{G}^\alpha ~ \text{on}~\mathcal{I}_0, \label{epd8}
\end{equation}
where $\alpha \in \{\gamma, \eta\}$. 
There exist unique pseudoinverse tensors $\left(\mathbb{H}^\alpha\right)^{-1}$ and $\left(\mathbb{G}^\alpha\right)^{-1}$ such that $ \left(\mathbb{H}^\alpha\right)^{-1} \mathbb{H}^\alpha = {\mathbbm 1}^\alpha~\text{and}~\left(\mathbb{G}^\alpha\right)^{-1} \mathbb{G}^\alpha = {\mathbbm 1}$. Here and elsewhere, no summation is implied for repeated superscript $\alpha$ unless explicitly stated. The interfacial jacobians $j^\alpha = |({\boldsymbol G}^\pm)^{*} {\mathbb N}|$  measure the ratio of infinitesimal areas in the relaxed configuration with respect to the reference configuration. We note the following results for later application:
\begin{equation}
\left(\mathbb{G}^{\alpha}\right)^\ast = {j}^\alpha ({\mathbb N}^\alpha \otimes \mathbb{N}),~ \partial_{\mathbb{G}^\alpha} j^\alpha=j^\alpha (\mathbb{G}^\alpha)^{-T},~\partial_{\mathbb{N}^\alpha} j^\alpha = {\bf 0},~ \mathring{j}^\alpha =  j^\alpha \mathring{\mathbb{G}}^\alpha({\mathbb{G}^\alpha})^{-1} \cdot {\mathbbm 1}^\alpha, \label{epd12}
\end{equation}
for each $\alpha \in \{\gamma, \eta\}$. Similar relations hold for $\mathbb{F}$ and $\mathbb{H}$.

\subsection{Integral theorems}
\label{intthm}
We collect several integral theorems which will be useful in the following section to derive localized relations from global balance laws and dissipation inequality. Consider an arbitrary simply-connected region $\Omega \subset \mathcal{B}_0$ such that $S = \Omega \cap \mathcal{I}_0$ is the interface contained within $\Omega$. The boundary $\partial S$ of $S$ is a subset of the boundary $\partial \Omega$ of $\Omega$. For a piecewise differentiable field ${\boldsymbol a} \in \mathcal{V}$, defined in $\mathcal{B}_0$, the divergence theorem requires
\begin{equation}
\int_{\Omega} \Div {\boldsymbol a} \d V = \int_{\partial{\Omega}}{\boldsymbol a}\cdot{\boldsymbol N} \d A
- \int_{S}\llbracket {\boldsymbol a} \rrbracket \cdot {\mathbb N} \d A,
 \label{bdiv}
\end{equation}
where $\d V$ and $\d A$ denote the infinitesimal volume and area measures in $\mathcal{B}_0$, respectively. The field ${\boldsymbol N} \in \mathcal{V}$ is the unit normal to $\partial \Omega$.  Let ${\boldsymbol \nu} \in \mathcal{V}$ be the outward unit normal to the closed curve $\partial S$ such that 
 ${\mathbb N} \cdot {\boldsymbol \nu}=0$, i.e., ${\boldsymbol \nu}$ is tangential to $S$. For a continuously differentiable field ${\mathbbm v} \in \mathcal{V}$, defined over $\mathcal{I}_0$, such that 
${\mathbbm v} \cdot{\mathbb N}= 0$,
the surface divergence theorem yields \cite{basak15b}
\begin{equation}
\int_{S} \Div^S {\mathbbm v}  \d A = \int_{{\partial S}} {\mathbbm v} \cdot {\boldsymbol \nu} \d L,
 \label{idiv}
\end{equation}
where $\d L$ is the infinitesimal length measure over $\mathcal{I}_0$. The above results can be suitably modified for scalar and tensor fields.

Let $f \in \mathcal{R}$ be a piecewise continuous field in $\mathcal{B}_0$ and let $g \in \mathcal{R}$ be a continuously differentiable field over $\mathcal{I}_0$. The following transport relations hold \cite{basak15b}: 
\begin{equation} \begin{split}
& \frac{\d}{\d t}\int_{\Omega} f \d V = \int_{\Omega} \dot{f} \d V -
 \int_{S} \llbracket f \rrbracket U \d A~\text{and} \\
& \frac{\d}{\d t}\int_{S} g \d A 
= \int_{S}({\mathring g}-g\kappa U) \d A
+\int_{\partial{S}} g W \d L, \end{split}
\label{trans}
\end{equation}
where $W \in \mathcal{R}$ is the velocity of edge $\partial S$ along $\boldsymbol{\nu}$. These transport theorems can be suitably modified for vector and tensor fields.

\section{Balance laws and dissipation}
\label{sec:balance_laws}
In this section, we state the global balance laws associated with mass, nutrient, and momentum, and derive their local counterparts in the bulk, away from the interface, and on the interface. We also state the global form of the dissipation inequality and, after making constitutive assumptions on the nature of bulk and interfacial energies, arrive at local dissipation inequalities in the bulk and on the interface. The local inequalities are used to derive simple kinetic relations for the evolution of growth and interface migration. In particular, we emphasize the coupling between biochemistry and biomechanics in our growth model. Finally, as a brief digression, we use our framework to discuss the growth of a thin film over a growing substrate. 

\subsection{Mass balance}
\label{subsec:mass}
Considering sources of mass in the bulk $\Pi_B \in \mathcal{R}$ (per unit reference volume) and on the interface $\Pi_{S} \in \mathcal{R}$ (per unit reference area), the global mass balance for an arbitrary region $\Omega \subset \mathcal{B}_0$ can be expressed as
\begin{equation}
\label{eq:mass1}
 \frac{\d}{\d{t}} \left [ \int_{\Omega}  \rho_0 \d{V} +  \int_{S}  \delta_0 \d{A} \right] =   \int_{\Omega}  \Pi_B \d{V} + \int_{S} \Pi_S \d{A} + \int_{\partial S} \delta_0 W \d{L},
\end{equation}
where $\rho_0 \in \mathcal{R}^+$ is the bulk mass per unit reference volume and $\delta_0 \in \mathcal{R}^+$ is the interfacial mass per unit reference area. The latter should be understood as an excess thermodynamic field, in the manner of Gibbs, for the non-material interface $S$. The sources of mass as diffusive fluxes, across the boundaries of both the bulk and the interface, are ignored. This is reasonable since we will be working with only simple elastic solids and their incorporation would otherwise require a higher-gradient constitutive theory \cite{epstein2000}. The last term in  \eqref{eq:mass1} represents the mass flow across $\partial S$ due to a part of the interface $S$ entering/leaving the fixed domain $\Omega$. Using the transport theorems \eqref{trans}, and then localizing the resulting integral equation, we obtain the local mass balance equations
\begin{equation} \label{eq:mass3}
 \begin{split}
   & \dot{\rho}_0 = \Pi_B~ \text{in}~ \mathcal{B}_0/\mathcal{I}_0~\text{and} \\
   & (\mathring{\delta}_0 - \delta_0 \kappa U) = \llbracket \rho_0 \rrbracket U +  \Pi_S ~ \text{on} ~\mathcal{I}_0. \end{split}
\end{equation}
The bulk mass source $\Pi_B$ can be related to the evolution of growth distortion tensor $\boldsymbol{G}$. Indeed, assuming that the bulk mass density, per unit volume of the relaxed configuration, remains unchanged for a fixed material point, i.e.  $\dot{\rho}_i = 0$, where $\rho_0 = J_G \rho_i$, we obtain $ \rho_0 \tr (\dot{\boldsymbol G} \boldsymbol{G}^{-1}) = \Pi_B$ \cite{ambrosi02}. Under elastic incompressibility ($J_H = 1$) this is equivalent to assuming $\dot{\rho} = 0$, where $\rho$ is the bulk mass density per unit volume of the current configuration such that $\rho_0 = J_F \rho$ \cite{Rodriguez1994}. On the other hand, the interfacial mass source $\Pi_S$ is related to both the areal evolution of growth tensor and the flux of bulk mass across the moving interface. In order to show this we assume, analogous to the bulk assumption, that the interfacial mass density ${\delta}_i^\alpha = \delta_0 (j^\alpha)^{-1}$, per unit area of the surface $\alpha \in \{\gamma, \eta\}$ in the relaxed configuration, remains conserved, i.e., $\mathring{\delta}_i^\alpha - \delta_i^\alpha \kappa U  = \llbracket \rho_i \rrbracket U$. It is only when the interface is stationary, or if it is a material surface ($U=0$), that these equations reduce to $\dot{\delta}_i^\alpha = 0$, an assumption previously made by Ciarletta et. al. \cite{ciarletta12}. The required relation can be readily obtained, by combining the assumed conservation law with \eqref{eq:mass3}$_2$ and \eqref{epd12}, as
\begin{equation} \label{eq:mass7}
\delta_0 \mathring{\mathbb{G}}^\alpha (\mathbb{G}^\alpha)^{-1} \cdot {\mathbbm{1}}^\alpha  + \left( j^\alpha \llbracket \rho_i \rrbracket - \llbracket \rho_0 \rrbracket \right) U  =  \Pi_S.
\end{equation}
If the interface is stationary, or if it is a material surface, then $\delta_0 \dot{\mathbb{G}}^\alpha (\mathbb{G}^\alpha)^{-1} \cdot {\mathbbm{1}}^\alpha  = \Pi_S$, a relationship similar to its bulk counterpart. For an elastically incompressible material, $\rho_i$ can be replaced by $\rho$ in \eqref{eq:mass7}. On the other hand, for an interface with no excess mass distribution, i.e., $\delta_0 = 0$, $-\llbracket \rho_0 \rrbracket U =  \Pi_S$; the interfacial mass source then necessarily requires a density variation across a moving interface.

\subsection{Nutrient balance}
\label{subsec:nut}

We represent all the biochemical nutrient activity in our body in terms of two nutrient concentration fields: $C \in \mathcal{R}^+$ (per unit reference volume) in the bulk and $ \mathbb{C} \in \mathcal{R}^+$ (per unit reference area) at the interface. The flux of nutrients is denoted by $\boldsymbol{M}  \in \mathcal{V}$ in the bulk and $\mathbb{M} \in \mathcal{V}$ at the interface.  We write the global nutrient balance law for an arbitrary region $\Omega \subset \mathcal{B}_0$ in the form
\begin{equation}\label{eq:conc1}
\begin{split}
   &\frac{\d}{\d t} \left[ \int_{\Omega} C \d{V} +  \int_{S} \mathbb{C} \d{A} \right] + \int_{\partial \Omega} \boldsymbol{M} \cdot \boldsymbol{N} \d{A}  + \int_{\partial S} \mathbb{M} \cdot \boldsymbol{\nu} \d{L} \\
   & \hskip .1in  = \int_{\Omega} \boldsymbol{E}_0 \cdot  \dot{\boldsymbol G} \boldsymbol{G}^{-1} \d{V} + \int_{S} \sum_{\alpha \in \{\gamma, \eta\}} {\mathbb E}_0^\alpha \cdot \mathring{\mathbb{G}}^\alpha (\mathbb{G}^\alpha)^{-1} \d{A} + \int_{\partial S}  \mathbb{C} W \d{L},
   \end{split}
\end{equation}
where the first term on the right side of the equality is the bulk source of nutrient concentration arising from the evolving growth tensor; $\boldsymbol{E}_0 \in Lin$ characterizes the anisotropy in the absorption rate of the nutrients  \cite{ambrosi07}. The second integral has analogous source terms for the interface characterized by ${\mathbb{E}}_0^\gamma \in Lin$ and ${\mathbb{E}}_0^\eta \in Lin$. The form of the nutrient source terms is motivated from the dissipation rates appearing in the local dissipation inequalities derived in Section \ref{subsec:dissip}. 
The last term in  \eqref{eq:conc1} represents nutrient flow across $\partial S$ due to a part of the interface $S$ entering/leaving the fixed domain $\Omega$. The global balance in \eqref{eq:conc1} can be localized, using transport and divergence theorems from Section \ref{intthm}, to obtain
 \begin{equation}
 \label{eq:conc2}
 \begin{split}
& \dot{C} + \Div \boldsymbol{M} = {\boldsymbol{E}}_0 \cdot  \dot{\boldsymbol{G}} \boldsymbol{G}^{-1} ~ \text{in}~\mathcal{B}_0/\mathcal{I}_0~\text{and} \\
   &       (\mathring{\mathbb{C}} - \kappa \mathbb{C} U)  + \Div^S{\mathbb{M}} -  \llbracket C \rrbracket U   +  \llbracket \boldsymbol{M} \rrbracket \cdot \mathbb{N} =  \sum_{\alpha \in \{\gamma, \eta\}} {\mathbb{E}}_0^\alpha \cdot \mathring{\mathbb{G}}^\alpha (\mathbb{G}^\alpha)^{-1} ~ \text{on} ~\mathcal{I}_0. \end{split}
\end{equation}
The nutrient balance laws relate biochemistry of the nutrients to biological growth \cite{ambrosi07,buskohl14,oller10}. The right hand sides therein couple nutrient concentration evolution to the growth evolution which are in turn governed by kinetics laws such as those obtained in Section \ref{subsec:dissip}. The interfacial concentration evolution is also influenced by the migration of the interface, which is governed by a kinetic law derived  in Section \ref{subsec:dissip}. The balance law \eqref{eq:conc2}$_1$ was first obtained by  Ambrosi and Guillou \cite{ambrosi07}.

\subsection{Momentum balance}
\label{subsec:momen}
Let $\boldsymbol{P} \in Lin$ and $\mathbb{P} \in Lin$ denote the bulk and the interfacial first Piola-Kirchhoff stress, respectively such that $\mathbb{P} \mathbb{N} = \boldsymbol{0}$. For $\Omega \subset \mathcal{B}_0$, the linear momentum balance requires
\begin{equation}
\label{eq:mom_bal_1}
 \int_{\partial \Omega} \boldsymbol{P} \boldsymbol{N} \d{A} + \int_{\partial S} \mathbb{P} \boldsymbol{\nu} \d{L}  = \boldsymbol{0},
\end{equation}
where both inertia and body force contributions have been neglected. Using the divergence theorems from Section \ref{intthm} and localizing, the global balance reduces to \cite{Gupta12}
\begin{equation}
\label{eq:mom_bal_5}
 \Div \boldsymbol{P} = \boldsymbol{0} ~ \text{in}~ \mathcal{B}_0/\mathcal{I}_0 ~\text{and}~  \Div^S \mathbb{P} +  \llbracket \boldsymbol{P} \rrbracket \mathbb{N} = \boldsymbol{0}~ \text{on} ~\mathcal{I}_0.
\end{equation}
On the other hand, in the absence of bulk and interfacial couples, the angular momentum balance requires \cite{Gupta12}
\begin{equation}
\label{eq:ang_mom_8}
   \boldsymbol{P} \boldsymbol{F}^T = \boldsymbol{F} \boldsymbol{P}^T  ~ \text{in}~ \mathcal{B}_0/\mathcal{I}_0 ~\text{and}~
   \mathbb{P} \mathbb{F}^T = \mathbb{F} \mathbb{P}^T ~ \text{on} ~\mathcal{I}_0.
\end{equation}

\subsection{Dissipation inequality and kinetic laws}
\label{subsec:dissip}
Under isothermal conditions, the second law of thermodynamics requires that the rate of change of the total free energy must be less than or equal to the mechanical power input. Denoting $\Psi_B \in \mathcal{R}$ and $\Psi_S \in \mathcal{R}$ as the bulk free energy (per unit reference volume) and the excess interfacial free energy (per unit reference area), respectively, we write the mechanical version of the second law of thermodynamics, neglecting inertia and body forces, for $\Omega \subset \mathcal{B}_0$ as
\begin{equation}
\label{eq:dissp1}
 \begin{split}
& \underbrace{\frac{\text{d}}{\text{d}t} \left( \int_{\Omega} \Psi_B \d{V} + \int_S  \Psi_S \d{A} \right ) }_{\text{rate of change of total free energy}} \leq
 \underbrace{\int_{\partial \Omega} \boldsymbol{P} \boldsymbol{N} \cdot \boldsymbol{v} \d{A} + \int_{\partial S} \mathbb{P} \boldsymbol{\nu} \cdot \mathbbm{v} \d{L} }_{\text{mechanical power input}} \\
&   - \underbrace{\int_{\partial \Omega}  \mu \boldsymbol{M} \cdot \boldsymbol{N} \d{A} - \int_{\partial S}  \mu \mathbb{M} \cdot \boldsymbol{\nu} \d{L}}_{\text{power due to nutrient flux}} + \underbrace{ \int_{\partial S} \mathfrak{C} \cdot \mathbbm{w} \d{L} + \int_{\partial S} \mathbb{P} \boldsymbol{\nu} \cdot \mathbbm{v}^{ext} \d{L} - \int_{\partial S}  \mu \mathbb{C} W \d{L}.}_{\text{non-standard power}}
 \end{split}
\end{equation}
In writing the above relation, owing to chemical equilibrium, we have assumed the interfacial chemical potential to be identical with either of the limiting  values of the bulk chemical potential $\mu$. The first two integrals on the right hand side of the inequality in \eqref{eq:dissp1} are power inputs due to bulk and interfacial tractions acting on $\partial \Omega$ and $\partial S$, respectively. The third and fourth integrals are entropic contributions due to nutrient fluxes. An alternative viewpoint is to consider the entropies directly associated with the incoming mass \cite{epstein2000, ciarletta12}. Following Ambrosi and Guillou \cite{ambrosi07}, we choose to work with the nutrients since we view growth to be an outcome of biochemical synthesis. The last three integrals are non-standard. The first of these is to account for excess entropy generation due to a part of the interface $S$ entering/leaving the fixed domain $\Omega$ \cite{basak15a,basak15b}, where $\mathbbm{w} \in \mathcal{V}$ is the intrinsic velocity of the edge $\partial S$, such that $\mathbbm{w}=U \mathbb{N}+W \boldsymbol{\nu}$. The second one provides a correction to the mechanical power due to interfacial traction. Indeed, the intrinsic material velocity $\mathbbm{v}$ shifts the observer, sitting at a point on the interface, away from $\partial \Omega$ while the extrinsic
material velocity on $\partial S$, $\mathbbm{v}^{ext} = W \mathbb{F} \boldsymbol{\nu}$, brings her back to the edge on $\partial \Omega$ \cite{basak15a}. The third one represents the excess entropy contribution from the nutrient flux as a part of the interface $S$ enters/leaves the fixed domain $\Omega$; see the last term in \eqref{eq:conc1}.
 The exact form of the non-standard force $\mathfrak{C} \in \mathcal{V}$ depends on the constitutive form of the interfacial energy, interfacial stress, and the dissipative fluxes. 
Towards this end, we assume the free energy densities to depend on elastic distortion and nutrient concentration:
\begin{equation}\label{eq:loc_dissp_1}
   \Psi_B=J_G \widetilde{\Psi}_B (\boldsymbol{H}, C)~ \text{and}~ \Psi_S=j^{\gamma} \widetilde{\Psi}_S (\mathbb{H}^{\gamma}, \mathbb{H}^{\eta}, \mathbb{C}),
\end{equation}
where $\widetilde{\Psi}_B$ is the free energy per unit volume of the bulk in the stress-free configuration and $\widetilde{\Psi}_S$ is the free energy per unit area of the $\gamma$-surface  in the relaxed configuration \cite{Gupta12}. 

The global relation in \eqref{eq:dissp1} can be localized with the help of divergence and transport theorems from Section \ref{intthm}, and further simplified using the local balance laws derived in the preceding sections. Localizing in the bulk, away from the interface, we use the standard arguments to obtain the constitutive relations
\begin{equation}
\label{eq:loc_dissp_5}
 \boldsymbol{P}= J_G \partial_{\boldsymbol{H}} \widetilde{\Psi}_B \boldsymbol{G}^{-T} ~ \text{and} ~ \mu= J_G \partial_C \widetilde{\Psi}_B ~ \text{in}~ \mathcal{B}_0/\mathcal{I}_0,
\end{equation}
and the local dissipation inequality \cite{ambrosi07}
\begin{equation} \label{eq:loc_dissp_6}
  (\widetilde{\boldsymbol{E}} + \mu \boldsymbol{E}_0) \cdot \dot{\boldsymbol{G}} \boldsymbol{G}^{-1}   + \Grad \mu \cdot \boldsymbol{M}   \leq 0 ~ \text{in}~ \mathcal{B}_0/\mathcal{I}_0, 
\end{equation}
where $\widetilde{\boldsymbol{E}}=J_G\left(\widetilde{\Psi}_B \boldsymbol{1}- \boldsymbol{H}^T\partial_{\boldsymbol{H}} \widetilde{\Psi}_B \right)$ is the elastic Eshelby tensor in the bulk \cite{Guptaetal07}. The local relations on the interface can be obtained by making note of the following identities:
\begin{eqnarray}
& \llbracket \boldsymbol{P}^T \boldsymbol{v} \rrbracket \cdot \mathbb{N} = -U \llbracket \boldsymbol{F}^T \boldsymbol{P} \rrbracket \mathbb{N} \cdot \mathbb{N} - \Div^S \mathbb{P} \cdot \mathbbm{v};\\
& \Div^S (\mathbb{P}^T \mathbbm{v}) = \Div^S \mathbb{P} \cdot \mathbbm{v} + \mathbb{P} \cdot \mathring{\mathbb{F}} - U \mathbb{F}^T \mathbb{P} \cdot \mathbb{L}; \\
& \mathring{\Psi}_S = \Psi_S \tr \left(\mathring{\mathbb{G}}^\gamma (\mathbb{G}^\gamma)^{-1}\right) + j^\gamma \left( \displaystyle{\sum_{\alpha \in \{\gamma, \eta\}}} \partial_{\mathbb{H}^\alpha} \widetilde{\Psi}_S \cdot \mathring{\mathbb{H}}^\alpha \right) + j^\gamma \partial_{\mathbb{C}} \widetilde{\Psi}_S \cdot \mathring{\mathbb{C}}.
\end{eqnarray}
Using standard arguments \cite{Gupta12}, we can obtain the constitutive relations
\begin{equation}
\label{eq:loc_dissp_12}
 \mathbb{P}= j^\gamma  \sum_{\alpha \in \{\gamma, \eta\}} \partial_{\mathbb{H}^\alpha} \widetilde{\Psi}_S (\mathbb{G}^\alpha)^{-T} ~ \text{and} ~ \mu= j^\gamma \partial_{\mathbb{C}} \widetilde{\Psi}_S~\text{on}~\mathcal{I}_0,
\end{equation}
and the dissipation inequality
\begin{equation}
\label{eq:loc_dissp_13}
\sum_{\alpha \in \{\gamma, \eta\}}  (\widetilde{\mathbb{E}}^\alpha + \mu {\mathbb{E}}_0^\alpha)   \cdot \mathring{\mathbb{G}}^\alpha(\mathbb{G}^\alpha)^{-1}  -  {f} U  +   \Grad^S \mu \cdot \mathbb{M}   \leq 0 ~\text{on}~\mathcal{I}_0,
\end{equation}
where $\widetilde{\mathbb{E}}^\gamma=j^\gamma (\widetilde{\Psi}_S \mathbbm{1}- (\mathbb{H}^\gamma)^T \partial_{\mathbb{H}^\gamma} \widetilde{\Psi}_S)$ and $\widetilde{\mathbb{E}}^\eta=- j^\gamma (\mathbb{H}^\eta)^T \partial_{\mathbb{H}^\eta} \widetilde{\Psi}_S$ are the elastic interfacial Eshelby tensors, and ${f}$ is the driving force for the normal motion of the interface, given by \cite{Gupta12}
\begin{equation}\label{eq:loc_dissp_13a}
{f}=  \mathbb{N} \cdot \llbracket  {\boldsymbol{E}} \rrbracket \mathbb{N} + \mathbb{E} \cdot \mathbb{L}.
\end{equation}
Here $\boldsymbol{E} = ({\Psi}_B + \mu C) \boldsymbol{1} - \boldsymbol{F}^T \boldsymbol{P}$ and $\mathbb{E}=(\Psi_S + \mu \mathbb{C}) \mathbbm{1}- \mathbb{F}^T \mathbb{P}$ are bulk and interfacial Eshelby tensors, respectively; note the difference between these Eshelby tensors with their elastic counterparts defined above. Finally, collecting all the leftover terms within the line integral over $\partial S$, and requiring that there is no excess entropy production at the edge, we obtain a constitutive representation for $\mathfrak{C}$:
\begin{equation}\label{eq:loc_dissp_18}
   \mathfrak{C} = \mathbb{E} \boldsymbol{\nu}.
\end{equation}
It represents the configurational force at the edge $\partial S$ of the interface as it propagates through the body. It should be noticed that if $\partial S$ represents an actual physical edge or a corner, for instance a kink in the interface, and not just an arbitrary domain, as considered above, then the non-standard power terms would no longer be needed in \eqref{eq:dissp1}.

The bulk dissipation inequality \eqref{eq:loc_dissp_6}  is identically satisfied if the following decoupled kinetic laws are assumed  \cite{ambrosi07}:
\begin{equation}
\label{eq:loc_dissp_8}
 \dot{\boldsymbol{G}} \boldsymbol{G}^{-1} = -g(C) (\widetilde{\boldsymbol{E}} + \mu \boldsymbol{E}_0)  ~\text{and}~  \boldsymbol{M} = -\boldsymbol{K}_0 \Grad \mu ~ \text{in}~ \mathcal{B}_0/\mathcal{I}_0,
\end{equation}
where $g \in \mathcal{R}^+$ and $\boldsymbol{K}_0 \in Lin$ is positive-definite. For positive mass addition $\tr(\widetilde{\boldsymbol{E}} + \mu \boldsymbol{E}_0) < 0 $, and vice-versa. Similarly, the interfacial dissipation inequality \eqref{eq:loc_dissp_13} is identically satisfied if the follwing decoupled kinetic laws are assumed on the interface:
\begin{equation}
\label{eq:loc_dissp_15}
\begin{split}
& \mathring{\mathbb{G}}^\gamma(\mathbb{G}^\gamma)^{-1} = -h_1(\mathbb{C}) \left(\widetilde{\mathbb{E}}^\gamma + \mu {\mathbb{E}}_0^\gamma \right),~ \mathring{\mathbb{G}}^\eta(\mathbb{G}^\eta)^{-1} = -h_2(\mathbb{C}) \left(\widetilde{\mathbb{E}}^\eta + \mu {\mathbb{E}}_0^\eta \right),\\
   &   U = {M} {f},~ \text{and}~ \mathbb{M} = - \mathbb{K}_0 \Grad^S \mu ~ \text{on}~ \mathcal{I}_0,
   \end{split}
\end{equation}
where $h_1 \in \mathcal{R}^+$, $h_2 \in \mathcal{R}^+$, ${M} \in \mathcal{R}^+$, and  $\mathbb{K}_0 \in Lin$ is positive-definite. 
It is clear from the growth evolution laws in the above kinetic relations that growth is possible as a result of both mechanical stresses, through the dependence on Eshelby tensors, and due to nutrient fluxes. Reciprocally, it is evident from \eqref{eq:conc2}, after substitutions from the above kinetic laws, that the nutrient concentration evolution is governed by stresses, nutrient fluxes, and interface migration. The complete initial-boundary-value problem, for determining the deformation, growth, and concentration fields, consists of Equations \eqref{eq:conc2},  \eqref{eq:mom_bal_5}, \eqref{eq:ang_mom_8}, \eqref{eq:loc_dissp_5}, \eqref{eq:loc_dissp_12}, \eqref{eq:loc_dissp_8}, and \eqref{eq:loc_dissp_15}, supplemented by initial conditions for concentration and growth distortion fields, and appropriate boundary data.

\subsection{Growing thin film over a growing substrate}
\label{filmonsub}
\begin{figure}[t!]
 \centering \includegraphics[width=0.35\textwidth, angle=270]{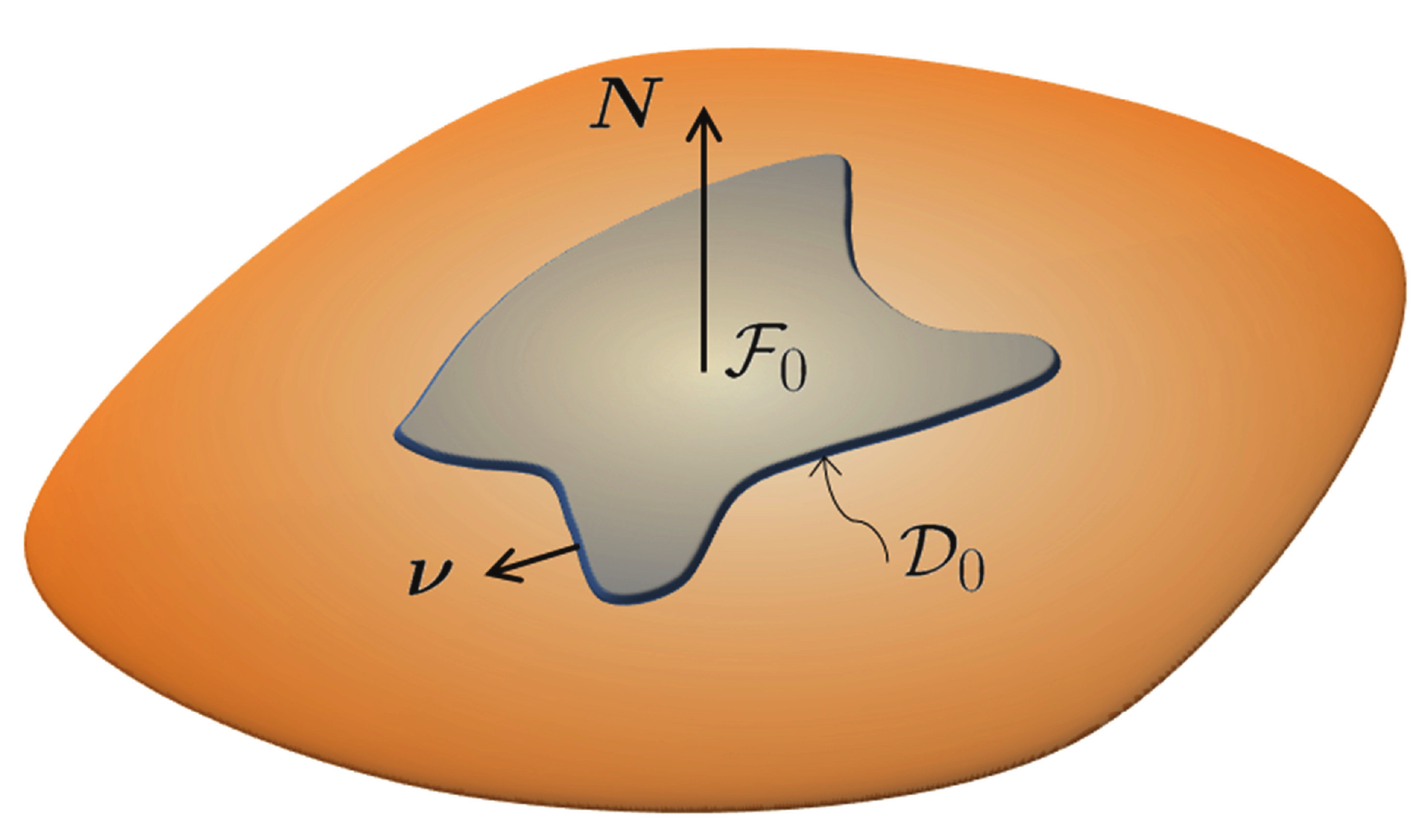}
  \caption{A growing thin film $\mathcal{F}_0$, bounded by a closed curve $\mathcal{D}_0$, over a growing substrate.}
  \label{film}
\end{figure}
Our framework can be used, with minor modifications, to develop a theory of growing elastic films bonded to growing elastic substrates. Such a formulation has been recently proposed by Kuhl and coauthors \cite{kuhl13a, kuhl13b} to model a variety of surface growth phenomena in biological systems. Our intent in the following, as a brief digression, is to recover their results while extending them to include biochemistry, more general kinetic laws, and boundary conditions at the film edge.  The interface of the preceding discussion now exists between a three-dimensional bulk solid and a two-dimensional thin film, see Figure \ref{film}. For simplicity, we will assume the interface energy, interfacial stress, and interfacial mass density to vanish identically. Let the thin film domain be denoted by $\mathcal{F}_0$ in the reference configuration. Its motion coincides with that of the bulk domain restricted to the interface. The surface deformation gradient field over $\mathcal{F}_0$ is defined as $\fixwidehat{\boldsymbol{F}} = \Grad^f \boldsymbol{\chi}$, where $\Grad^f$ represents the surface gradient. We have $\fixwidehat{\boldsymbol{F}} = \boldsymbol{F}^- \mathbbm{1}$, where $\boldsymbol{F}^-$ is the limiting value of the deformation gradient in the bulk as it approaches the interface, due to coherency of the total deformation; $\mathbbm{1} = \boldsymbol{1} - \boldsymbol{N} \otimes \boldsymbol{N}$ is the projection tensor associated with $\mathcal{F}_0$, where $\boldsymbol{N} \in \mathcal{V}$ is the unit normal field on $\mathcal{F}_0$. The surface deformation gradient admits a multiplicative decomposition, analogous to the bulk, as $\fixwidehat{\boldsymbol{F}} = \fixwidehat{\boldsymbol{H}} \fixwidehat{\boldsymbol{G}}$, where $\fixwidehat{\boldsymbol{H}} \in Lin$ and $\fixwidehat{\boldsymbol{G}} \in Lin$ are, respectively, elastic and growth distortion tensor fields over $\mathcal{F}_0$. The interface between the bulk substrate and the thin film is, in general, incoherent and therefore neither $\fixwidehat{\boldsymbol{H}}$ nor $\fixwidehat{\boldsymbol{G}}$ are projections of their bulk counterparts. 

The local governing equations for the substrate remain same as those derived for the bulk in the preceding sections. The local mass balance for the film requires $\dot{\hat \delta} = \Pi_f$, where $\hat{\delta} \in \mathcal{R}^+$ is mass per unit reference area of the thin film and $\Pi_f \in \mathcal{R}$ is the corresponding mass source. Furthermore, if $\fixwidehat{C} \in \mathcal{R}^+$ is the nutrient concentration (per unit reference area) and $\fixwidehat{\boldsymbol{M}} \in \mathcal{V}$ is the nutrient flux field over $\mathcal{F}_0$, the nutrient balance for the thin film is of the form $({\fixwidehat C})\dot{} + \Div^f \fixwidehat{\boldsymbol{M}} - \boldsymbol{M} \cdot \boldsymbol{N}= {\hat{\boldsymbol{E}}}_0 \cdot  ({\fixwidehat{\boldsymbol{G}}})\dot{} {\fixwidehat{\boldsymbol{G}}}^{-1}$, where ${\fixwidehat{\boldsymbol{E}}}_0$ is a constant tensor and ${\fixwidehat{\boldsymbol{G}}}^{-1}$ is the pseudoinverse of ${\fixwidehat{\boldsymbol{G}}}$. The momentum balances in the thin film region require $\Div^f \fixwidehat{\boldsymbol P} - \boldsymbol{P} \boldsymbol{N} = \boldsymbol{0}$ and $ \fixwidehat{\boldsymbol P} \fixwidehat{\boldsymbol F}^T =  \fixwidehat{\boldsymbol F} \fixwidehat{\boldsymbol P}^T$, where $\fixwidehat{\boldsymbol P} \in Lin$ is the surface first Piola-Kirchhoff stress on $\mathcal{F}_0$. We consider, treating $\mathcal{F}_0$ as an hyperelastic membrane, a free energy density per unit area of the stress-free configuration as $\fixwidehat{\Psi} (\fixwidehat{\boldsymbol H}, \fixwidehat{C})$.  It is then straightforward to employ the dissipation inequality for the material points occupying $\mathcal{F}_0$ to obtain, on one hand, $\fixwidehat{\boldsymbol P} = \hat{j} \partial_{\fixwidehat{\boldsymbol{H}}} \fixwidehat{\Psi} \fixwidehat{\boldsymbol{G}}^{-T}$  and $\mu= \hat{j} \partial_{\fixwidehat{C}} \fixwidehat{\Psi}$ and, on the other, $(\fixwidehat{\boldsymbol{E}} + \mu \fixwidehat{\boldsymbol{E}}_0) \cdot ({\fixwidehat{\boldsymbol{G}}})\dot{} \fixwidehat{\boldsymbol{G}}^{-1}   + \Grad^f \mu \cdot \fixwidehat{\boldsymbol{M}}  \leq 0$, such that $\hat{j} \in \mathcal{R}^+$ is the ratio of infinitesimal areas of the film in the stress-free configuration with respect to the reference configuration and $\fixwidehat{\boldsymbol{E}} =\hat{j} \left(\fixwidehat{\Psi} \mathbbm{1}- \fixwidehat{\boldsymbol{H}}^T\partial_{\fixwidehat{\boldsymbol{H}}} \fixwidehat{\Psi} \right)$ is the elastic surface Eshelby tensor; compare these with \eqref{eq:loc_dissp_5}-\eqref{eq:loc_dissp_6} and \eqref{eq:loc_dissp_12}-\eqref{eq:loc_dissp_13}. The kinetic laws which satisfy the inequality are
\begin{equation}
\label{eq:kinthinfilm}
 ({\fixwidehat{\boldsymbol{G}}})\dot{} \fixwidehat{\boldsymbol{G}}^{-1} = -\hat{h}(\fixwidehat{C}) (\fixwidehat{\boldsymbol{E}} + \mu \fixwidehat{\boldsymbol{E}}_0)  ~\text{and}~  \fixwidehat{\boldsymbol{M}} = -\fixwidehat{\boldsymbol{K}} \Grad^f \mu ~ \text{in}~ \mathcal{F}_0,
\end{equation}
where $\hat{h} \in \mathcal{R}^+$ and $\fixwidehat{\boldsymbol{K}} \in Lin$ is positive-definite. These can be substituted back into the equations of nutrient mass balance to deduce the evolution equations for nutrient concentration over the thin film. These equations also act as the boundary conditions for the differential equations which govern the nutrient concentration in the substrate. Additionally, the following boundary conditions at the film edge $\mathcal{D}_0$ (see Figure \ref{film}), in terms of a prescribed nutrient flux $\fixwidehat{m}\in \mathcal{R}$ and traction $\fixwidehat{\boldsymbol{t}} \in \mathcal{V}$, need to be satisfied:
\begin{equation}
\fixwidehat{M} \cdot \boldsymbol{\nu} = \fixwidehat{m},~\text{and}~ \fixwidehat{\boldsymbol{P}} \boldsymbol{\nu} - \lim_{\epsilon\rightarrow 0} \oint_{C_\epsilon} \boldsymbol{PN} \d L  = \fixwidehat{\boldsymbol{t}}, ~ \text{on}~ \mathcal{D}_0, \label{edge}
\end{equation}
where $C_\epsilon$ is the boundary of a small semi-circular disc of radius $\epsilon$ centred at a point on $\mathcal{D}_0$ \cite{basak15a,basak15b}. In writing \eqref{edge}$_1$, we assume the bulk concentration field $C$ to remain bounded at the film edge. The film edge will not contribute to dissipation as long as there is no intrinsic nutrient flux, stress, or energy associated with it. 

\section{Tree growth due to ring formation}
\label{sec:rings}
Trees increase their girth by forming a new ring of wood over the existing trunk \cite{plomion01}. The deposition of wood takes place in a thin layer of Xylem and Phloem cells, known as vascular cambium, between the still developing ring and the bark \cite{pallardy08}. Our interest is to model the emergence of growth stresses and nutrient concentration field in trees due to mass addition in this thin layer. The growth stresses are different from the stresses which are generated in response to mechanical loading (e.g., due to wind) or those which appear due to sharp changes in the moisture content of the tree \cite{archer86}.  The cell swelling is understood to induce compressive growth stresses along the periphery of the trunk, whereas the longitudinal shortening develops tensile growth stresses along the length of the trunk \cite{wilkins86}. The stresses generated in a new layer cumulatively bring about stress gradients in the overall structure such that the longitudinal stresses are tensile on the outer surface of the trunk and compressive at the center, while the circumferential and radial stresses are compressive on the outside and tensile at the center \cite{plomion01, archer86, mattheck97}. On the other hand, the nutrient activity is restricted to a small neighborhood of the vascular cambium interface including the recently formed ring and a portion of the bark. The nutrient concentration is maximum at the interface and decreases steadily into the ring and the bark domains.

The growth stresses have been calculated previously  \cite{archer86,fournier90} by combinining bulk growth with an incremental approach, where elasticity of the bark, non-uniformity in the ring sizes, and nutrient fields were all neglected.  The growth strains were estimated either by relieving stresses from the outermost ring at each increment \cite{archer86} or using the microstructure data \cite{fournier90}. The latter method was in fact devised to replace the former which did not yield actual growth strains in the inner rings of the tree. We revisit the problem in the framework of incoherent interfacial growth with nutrient driven mass addition at the vascular cambium interface. Moreover, we provide a novel way to estimate the growth strain field in the trunk by exploiting the non-uniform ring size distribution and using the available experimental data for the growth strains in the bark and the latest ring. Biologically, the growth strains are directly dependent on the amount of lignin and cellulose deposited in the cells during wood formation \cite{sugiyama93}, and should therefore be directly related to the relative size, or equivalently the mass (assuming a constant density for the wood), of the growing ring. 

\subsection{The model}
\begin{figure}[t]
 \centering \includegraphics[width=0.6\textwidth]{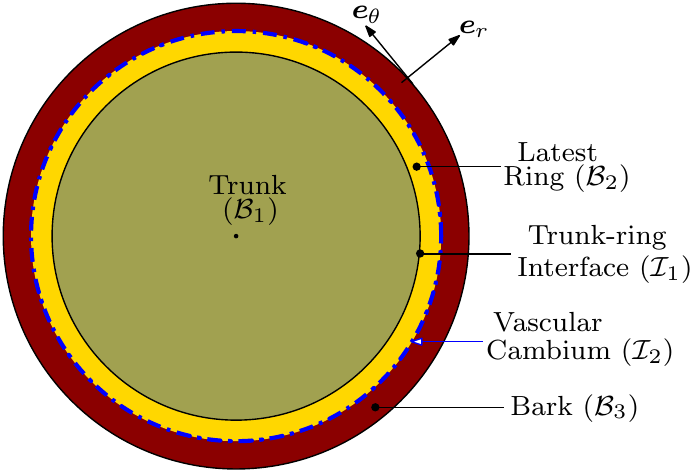}\\
  \caption{The tree growth model with three bulk domains separated by two interfaces. The axial direction $\boldsymbol{e}_z$ is normal to the plane of the paper.}
  \label{fig:3tier_model}
\end{figure}
We consider two interfaces in our model of tree growth: an interface $\mathcal{I}_1$ between the tree trunk domain (denoted by $\mathcal{B}_1$) and the recently formed ring (denoted by $\mathcal{B}_2$), and an interface $\mathcal{I}_2$ between $\mathcal{B}_2$ and the bark (denoted by $\mathcal{B}_3$), see Figure \ref{fig:3tier_model}. The combined configuration of the tree trunk, which includes both the pith and the matured rings, the latest ring, and the bark forms a long cylinder with a circular cross-section such that axisymmetry is maintained throughout. The two interfaces are oriented such that the associated normals point outward towards the bark. The mass addition, which takes place only at the vascular cambium interface $\mathcal{I}_2$, is responsible for both the formation of the new ring and the increase in girth of the bark. Accordingly, we decompose the mass source $\Pi_S$ into a component $\Pi_S^{\eta}$, responsible for the ring formation, and $\Pi_S^{\gamma}$, which contributes to the bark growth. The nutrient flux is also assumed to exist only at $\mathcal{I}_2$. 

Following earlier treatments \cite{archer86, fournier90},  we work with linearized strain kinematics such that $\boldsymbol{F} \approx \boldsymbol{1} + \boldsymbol{f}$, $\boldsymbol{H} \approx \boldsymbol{1} + \boldsymbol{h}$, and $\boldsymbol{G} \approx \boldsymbol{1} + \boldsymbol{g}$, all small to the same order. The multiplicative decomposition of the deformation gradient is hence replaced by the additive decomposition $\boldsymbol{f}=\boldsymbol{g}+\boldsymbol{h}$. The growth distortion field in the trunk domain is taken of the form $\boldsymbol{g}_1 (r,t)=k_{1r} (r,t) \boldsymbol{e}_r \otimes \boldsymbol{e}_r +  k_{1 \theta} (r,t) \boldsymbol{e}_\theta \otimes \boldsymbol{e}_\theta + k_{1 z} (r,t) \boldsymbol{e}_z \otimes \boldsymbol{e}_z$, where $r \in \mathcal{R}^+$ is the radial coordinate. On the other hand, the growth distortions in the newly formed ring and the bark are assumed to be spatially uniform as $\boldsymbol{g}_2(t)=k_{2r}(t) \boldsymbol{e}_r \otimes \boldsymbol{e}_r +  k_{2 \theta} (t) \boldsymbol{e}_\theta \otimes \boldsymbol{e}_\theta + k_{2z} (t) \boldsymbol{e}_z \otimes \boldsymbol{e}_z$ and $\boldsymbol{g}_3(t)=k_{3r}(t) \boldsymbol{e}_r \otimes \boldsymbol{e}_r +  k_{3\theta} (t) \boldsymbol{e}_\theta \otimes \boldsymbol{e}_\theta + k_{3z} (t) \boldsymbol{e}_z \otimes \boldsymbol{e}_z$, respectively. Using relations from Section \ref{subsec:mass}, we immediately obtain $\dot{k}_{Ir} + \dot{k}_{I \theta}+\dot{k}_{Iz}=0$ for no mass addition in the bulk, where $I=1,2,$ and $3$, and $\delta_0(\dot{k}_{2 \theta}+\dot{k}_{2z}+\dot{k}_{3\theta}+\dot{k}_{3z})=\Pi_S^\eta + \Pi_S^\gamma$ on $\mathcal{I}_2$. In deriving the latter, we have ignored the normal speed of the interface considering it to be much slower than the growth rate process. The interfacial equations are identically satisfied if we assume $\dot{k}_{2\theta}+\dot{k}_{2z}=\Pi_\eta/\delta_0$ and $\dot{k}_{3 \theta}+\dot{k}_{3z}=\Pi_\gamma/\delta_0$. To simplify further, we take ${k}_{2r}=k_{2\theta}$ and ${k}_{3r}=k_{3\theta}$  \cite{archer86}. As a result, growth distortions in the ring and the bark regions are completely determined in terms of the interfacial mass source. The growth distortions in the trunk, on the other hand, will be obtained in Section \ref{randd} using the ring size distribution in the matured trunk.

The residual stresses in a growing body are generated due to the elastic deformations, which appear in order to yield a connected body in the grown configuration. For an analytically tractable framework, we assume bulk elastic strain energies to be decoupled from bulk chemical energies, assume interfacial elastic energies to be negligible for both the interfaces, and consider a linearized stress-strain constitutive form with orthotropic elastic constants. Moreover, we consider a displacement field of the form $\boldsymbol{u}=u(r) \boldsymbol{e}_r + w(z) \boldsymbol{e}_z$ and limit our attention to a fixed time instance. The non-trivial stress-strain relationships, in terms of cylindrical coordinates, are
\begin{equation}
\label{eq:formuln5}
\begin{split}
& \sigma_{rr}= C_{rr} \left(u'(r) - k_r \right) + C_{r \theta} \left( u/r - k_{\theta} \right)  + C_{rz} \left (w'(z) - k_z\right),  \\
& \sigma_{\theta \theta}= C_{\theta r} \left( u'(r) - k_r  \right) + C_{\theta \theta} \left( u/r - k_{\theta} \right)  + C_{\theta z} \left (w'(z) - k_z \right),~\text{and}  \\
& \sigma_{zz}= C_{zr} \left(u'(r) - k_r \right) + C_{z \theta} \left( u/r - k_{\theta} \right)  + C_{zz} \left (w'(z) - k_z  \right),
\end{split}
\end{equation}
where the orthotropic elastic constants are such that $C_{r \theta} = C_{\theta r}$, $C_{rz} = C_{zr}$, and $C_{z \theta} = C_{\theta z}$; the superscript prime denotes the derivative of the function with respect to its argument. The governing equations for displacements can be obtained by substituting these relations into the equilibrium equations. The boundary conditions include traction-free outer surface of the bark, continuity of the radial stress and the displacement vector at the trunk-ring and the ring-bark interface, finiteness of the radial displacement at the center of the trunk, and zero net force arising out of longitudinal residual stress distribution in the trunk, ring, and bark. The problem is analytically solved by fitting the trunk growth distortion field into a quadratic function of $r$. 

\begin{figure}[t!]
\centering
 \subfloat[]{\includegraphics[width=0.42\textwidth]{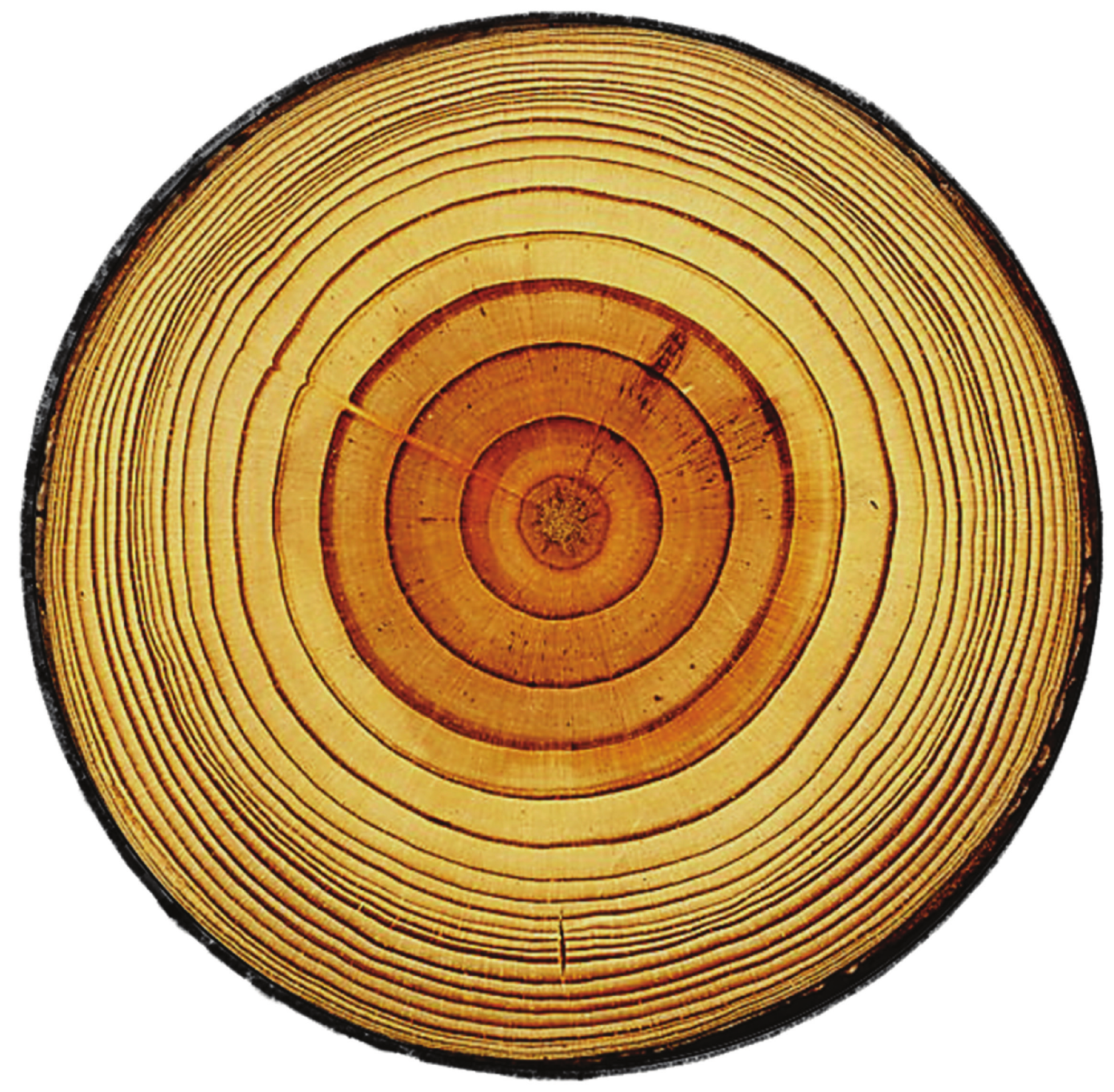}}
 \hspace{2mm}
 \subfloat[]{\includegraphics[width=0.55\textwidth]{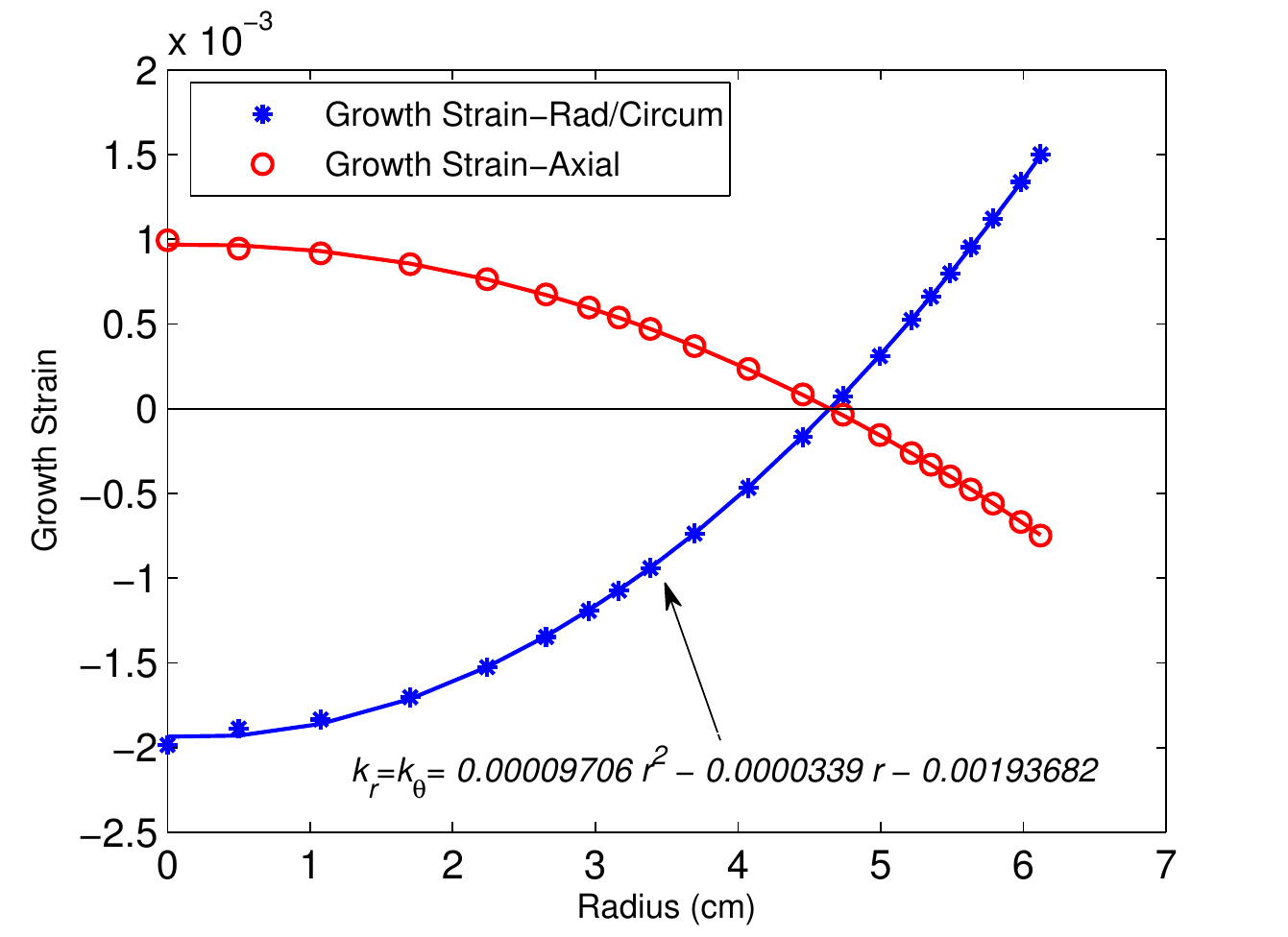}}
\caption{(a) Cross-section of a typical Pine tree trunk showing non-uniform distribution of rings. (b) The variation of the estimated growth strain field in the cross-section of the tree trunk.}
\label{trunkgs}
\end{figure}

\subsection{Results and discussion}
\label{randd}
 Assuming a uniform mass density of the wood, and using a calibration factor, we can convert the mass in each of the matured ring to a corresponding value of growth strain. Towards this end, we take the size distribution of the matured rings from a typical cross-section of a pine tree, shown in Figure \ref{trunkgs}(a). The strain in the outermost matured ring is estimated from the experimentally available value of the strain in the latest ring under analysis. A smooth curve is then fitted to obtain the non-uniform distribution of radial growth strain in the trunk domain, see Figure \ref{trunkgs}(b). We obtain $k_{1r}(r)=0.00009706 r^2 - 0.00003390 r - 0.00193682$. For the tangential and the axial growth strains we assume $k_{1\theta} = k_{1r}$ and $k_{1z}=-k_{1\theta}/2$ \cite{archer86,fournier90}.  The obtained distribution is in agreement with the trunk and plank stripping results of Archer and coauthors \cite{archer86, Post80}. 
The uniform growth strains in the latest ring and the bark are taken as $k_{2r} = k_{2\theta} = 0.002$, $k_{2z}=-k_{2\theta}/2$ and $k_{3r}=k_{3\theta}=-0.0002$, $k_{3z}=-0.0009$, respectively \cite{archer86,fournier90}. The value of the growth strains in the latest ring indicates that it has grown circumferentially, creating an overlap, and shortened axially from a hypothetical reference state of our model. Therefore, to obtain the connected final configuration of the body, we need compressive elastic strains in the $\theta$-direction and tensile elastic strains in the $z$-direction. Similar interpretations can be provided for growth strains in the trunk and the bark domains. The outer radius of the tree trunk is taken as $6.121$ cm for our calculations; the pith is assumed to be absent altogether. The thickness of the latest ring is taken as $0.053$ cm and of the bark as $0.25$ cm.
The orthotropic elastic constants for trunk, ring, and bark domains are taken to be identical as $C_{rr}=1560$ MPa, $C_{\theta \theta}=890$ MPa, $C_{zz}=12300$ MPa, $C_{r\theta}=620$ MPa, $C_{\theta z}=650$ MPa, and $C_{rz}=890$ MPa \cite{fournier90}.

\subsubsection{Growth stresses}
\label{gs}
The growth stresses obtained for the considered parametric values are shown in Figure \ref{fig:stress1}(a). The qualitative behaviour of stress fields in the trunk as well as the outermost ring domain is in good agreement with the existing literature \cite{archer86,fournier90}.  Also, as expected, the radial stress in the outermost ring as well in the bark remains vanishingly small. The circumferential stress is sharply discontinuous at both the interfaces.  Similarly, there is a sudden jump in the magnitude of the axial stress across the trunk-ring interface and again a smaller jump at the ring-bark interface. We repeated our calculation by varying the stiffness of the bark. Interestingly, decreasing the stiffness even by four times showed no significant influence on the stress values in the outer trunk and the latest ring region; the stresses in the bark, of course, vary significantly, as demonstrated in Table \ref{Table-1}(a). This can be understood by noting that due to force equilibrium in the axial direction, large stresses in the bark are compensated by smaller stresses in the trunk. The change in bark stiffness inversely affects the stress close to the center of the trunk. A decrease in bark stiffness hence makes it favorable for center cracking of the trunk, or in other words, a bark of sufficiently high stiffness would produce high quality timber. 

\subsubsection{Nutrient concentration during tree growth}
\begin{figure}[t!]
\centering
 \subfloat[]{\includegraphics[width=0.52\textwidth]{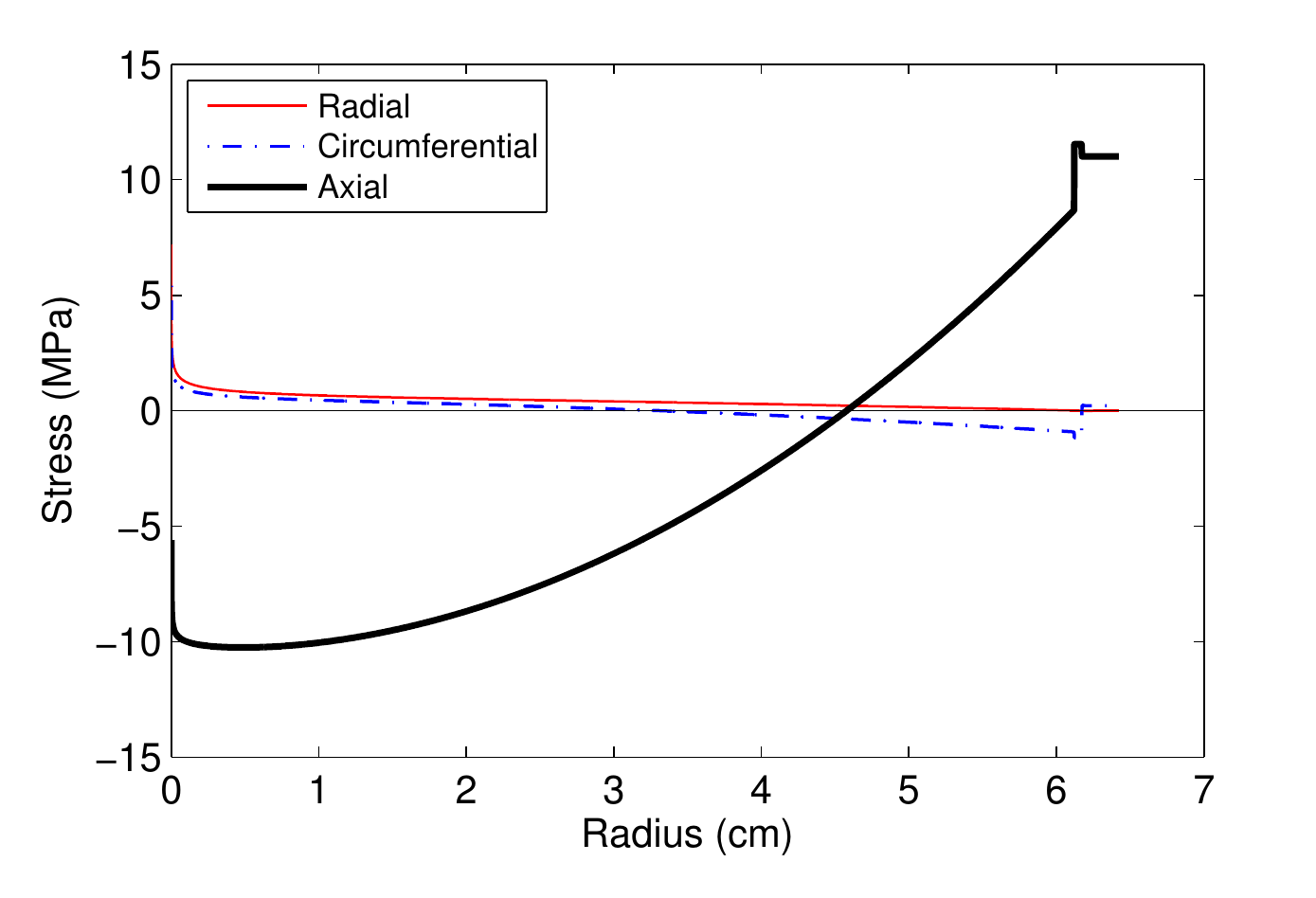}}
 \subfloat[]{\includegraphics[width=0.50\textwidth]{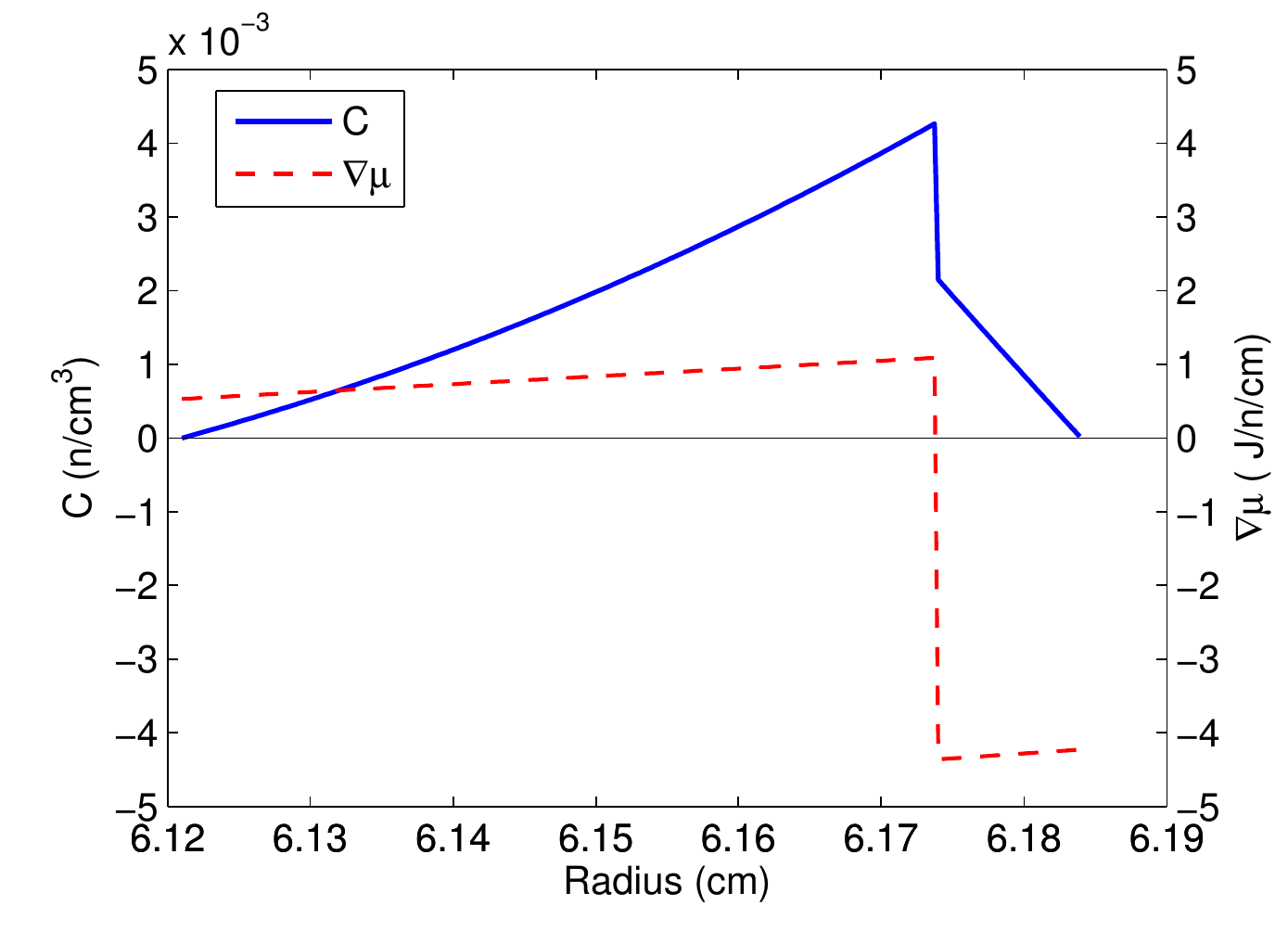}}
\caption{(a) Stress distribution in the Pine tree trunk. (b) The distribution of nutrient concentration and nutrient flux in the ring and bark domains; n$=10^8$ cells.}
\label{fig:stress1}
\end{figure}
The transportation of the nutrients, through the vascular cambium layer $\mathcal{I}_2$, assists in the proliferation of Xylem and Phloem cells in the recently formed ring $\mathcal{B}_2$ and in some portion of the bark $\mathcal{B}_3$, respectively. Considering steady state of the nutrient chemistry and a quasistatic deposition of the wood cells, i.e.,
$\dot{C}=0$, $\mathring{\mathbb{C}}=0$, $U=0$, we obtain the nutrient concentration field as a consequence of the growth and diffusion processes. The solution is meaningful at a time instance just before the maturation of the latest ring. We assume bulk concentration field to be such that it vanishes in $\mathcal{B}_1$ and varies only radially in $\mathcal{B}_2$ and $\mathcal{B}_3$. The interface concentration is assumed to vanish over $\mathcal{I}_1$ and to be constant over $\mathcal{I}_2$. The nutrient flux is assumed to be zero in $\mathcal{B}_1$ and over $\mathcal{I}_1$. The free energy density of the bulk regions are additively composed of a quadratic strain energy and a quadratic chemical energy term, such that $\mu = \alpha_a C$ (to the leading order in strain), where $\alpha_a = \alpha_2$ in $\mathcal{B}_2$ and $\alpha_a = \alpha_3$ in $\mathcal{B}_3$. The free energy density of the interface $\mathcal{I}_2$ consists only of a quadratic chemical energy term such that $\mu = \beta \mathbb{C}$ (to the leading order in strain). There is no mass addition in the bulk, leading to $\tr ({\dot{\boldsymbol{G}} \boldsymbol{G}^{-1}})=0$ or, equivalently, $\tr (\widetilde{\boldsymbol{E}} + \mu \boldsymbol{E}_0) = 0$. This is identically satisfied if we assume $ \mu \boldsymbol{E}_0 =-\tr(\widetilde{\boldsymbol{E}}) \boldsymbol{e}_r \otimes \boldsymbol{e}_r$. Chemical equilibrium at $\mathcal{I}_2$ require the limiting values of the bulk chemical potential, from either side of the interface, to be equal to the interfacial chemical potential; as a result, $ \alpha_3 C^+ =  \alpha_2 C^-$ and $ \alpha_2 C^- =\beta \mathbb{C}$. 
Finally, for analytical simplicity and purposes of computation, we choose $\boldsymbol{K}_0 = \boldsymbol{1}$ n$^2$/J-cm-s, $g(C) = C$, $\alpha_2 = 20$ Jcm$^3$/n$^2$, $\alpha_3 = 40$ Jcm$^3$/n$^2$, $\mathbb{E}_0 = -\mathbbm{1}$ n/cm$^2$, $\mathbb{K}_0 = \mathbbm{1}$ n$^2$/J-s, $\beta = 0.01$ Jcm$^2$/n$^2$, and $h_1(\mathbb{C})=h_2(\mathbb{C}) = 100\mathbb{C}$, where n denotes $10^8$ cells.
Under these conditions, the nutrient balance equations \eqref{eq:conc2}, with substitutions from kinetic laws \eqref{eq:loc_dissp_8} and \eqref{eq:loc_dissp_15}, reduce to (upto leading order in strain)
\begin{equation}
\label{trnteq1}
\alpha_a^2 \left( C''(r) + \frac{1}{r} C'(r) \right) = (\tr{\widetilde{\boldsymbol{E}}}) (\widetilde{{E}}_{\theta \theta} + \widetilde{{E}}_{zz}) ~\text{in}~\{\mathcal{B}_2 \cup \mathcal{B}_3 \} /\mathcal{I}_2 ~\text{and}
\end{equation}
\begin{equation}
\label{trnteq2}
   \left( \alpha_3 C'(r)^+ -  \alpha_2 C'(r)^- \right)  = \frac{100 \alpha_2 ^2 {C}^-}{\beta^2}  \left( - \alpha_2 (C^-)^2  + 4 \beta   {C^-}  \right) ~\text{on}~ \mathcal{I}_2,
\end{equation}
where $\widetilde{{E}}_{\theta \theta}$ and $\widetilde{{E}}_{zz}$ are the circumferential and axial components of the Eshelby tensor. Equation \eqref{trnteq1} is a second-order differential equation to be solved within the ring and bark domains. For boundary conditions, we assume the concentration to be zero both at the inner radius of the ring and at a radial distance of $0.01$ cm from $\mathcal{I}_2$ into the bark; the concentration is assumed to remain zero in rest of the bark. In addition, there are two interfacial conditions at $\mathcal{I}_2$ given by the continuity of the chemical potential and the interfacial nutrient balance \eqref{trnteq2}. The results are shown in \ref{fig:stress1}(b), where stresses from Figure \ref{fig:stress1}(a) have been used. As expected, the concentration is maximum at the vascular cambium interface and that it spreads more into the ring region than the bark. The latter is a representation of a larger spread of Xylem cells in comparison to Phloem cells.  The piecewise near constant behavior of the chemical potential gradient, on the other hand, indicates that the ring formation is in its final stage.

\subsubsection{Cracking pattern in the bark}
\begin{figure}[t!]
 \centering \includegraphics[width=0.45\textwidth, angle=270]{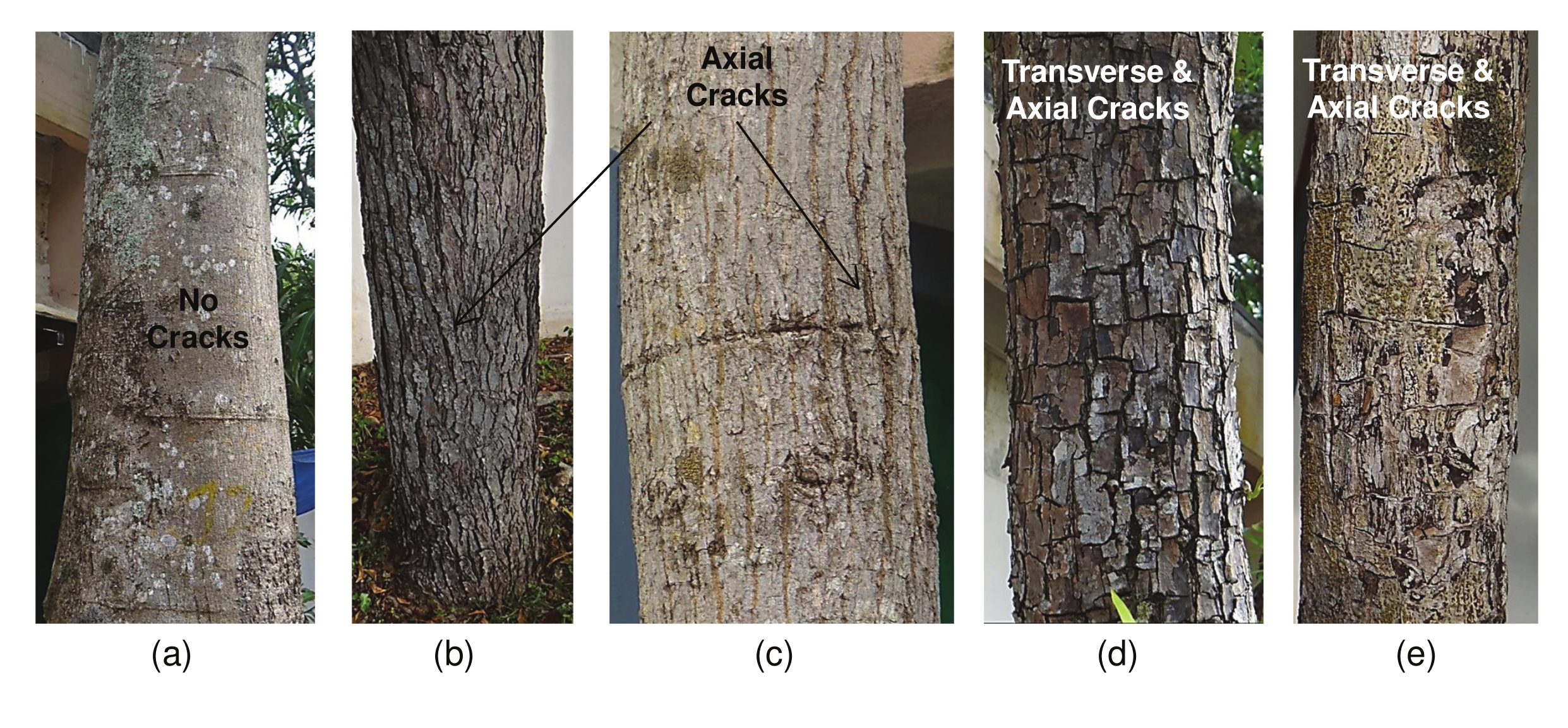} \vskip -0.2in
  \caption{Crack patterns on the outer bark of various Indian trees: (a) Delonix Regia (\emph{gulmohar}); (b) Azadirachta Indica (\emph{neem}); (c)  Magnifera Indica (\emph{mango}); (d) and (e) are both Terminalia Catappa (\emph{almond}) trees, located within $3$ m of each other, with (e) being the younger tree. An additional crack along the circumference in (c) is due to tying of a plastic rope over the tree for three years.}
  \label{fig:cracks}
\end{figure}
The outer bark in different trees cracks differently, as is shown for four common Indian trees in Figure \ref{fig:cracks}. In fact the bark in some trees does not crack, and in most of the trees the cracking depends on the age of the tree. We can use the visible crack pattern to infer qualitative details about the nature of growth strains and elastic moduli associated with the bark. Indeed, the stress value in the bark, which varies with growth strains and stiffness, can be correlated with the cracking pattern. In Table \ref{Table-1}(a) we report the variation in bark stresses for different bark stiffness values, while keeping all the other parameters as given in the beginning of Section \ref{randd}. These should be compared with the result in row (iii) of Table \ref{Table-1}(b), which corresponds to the case of equal elastic moduli in the bark and trunk regions. Both axial and circumferential stresses are higher for barks with increased stiffness. The severity of cracking in Figure \ref{fig:cracks}(d), compared to that in Figure \ref{fig:cracks}(e), can therefore be explained if we assume the bark stiffness to increase with the tree getting older.  
In Table~\ref{Table-1}(b) the bark stresses are compared for different combinations of growth strains in the bark region. Other parameters are kept fixed according to the values provided in the beginning of Section \ref{randd}. The first thing to note is that, whenever the growth strains are all compressive in nature (e.g., rows (iii) and (v)), tensile stresses are generated in both circumferential and axial directions. In fact, larger compressive strains lead to increased axial and circumferential tensile stresses. We can infer from the cracking patterns in Figures \ref{fig:cracks}(d) and \ref{fig:cracks}(e) that the bark growth strains therein are compressive in nature. Secondly, for sufficiently high axial tensile growth strains, in addition to sufficiently low compressive circumferential strains, we are led to purely compressive stress states in the tree bark; compare rows (iv) and (vii). Accordingly, we can correlate the absence of cracking in \ref{fig:cracks}(a) to such a situation. Thirdly, positive circumferential growth strains, combined with negative axial strains (e.g., row (vi)), can lead to a tensile axial stress but a compressive circumferential stress. This correlates to the formation of only transverse cracks in the bark. Finally, a case for only axial cracking in the bark, as for trees in Figures \ref{fig:cracks}(b) and \ref{fig:cracks}(c), can be made if one considers sufficiently high compressive circumferential growth strains coupled with low tensile axial growth strains (e.g., row (vii)). 

\begin{table}[t!]
  \centering
  \begin{tabular}{| l | l | l | l | l | l |}
    \hline
     Parameters & \multicolumn{3} {|c|}{Bark stresses in MPa} \\ \cline{2-4}
        & Radial  & Circumferential  & Axial   \\
    \hline \hline
     \multicolumn{4} {|c|}{(a) Variation in bark stiffness ($k_{3\theta}=k_{3r}=-0.0002$, $k_{3z}=-0.0009$)} \\
     \hline \hline
  (i) $\left(C_{ij}\right)_{\text{bark}} = 2 \left(C_{ij}\right)_{\text{trunk}}$   & 0 &	0.417 &	20.474\\
  \hline
(ii) $\left(C_{ij}\right)_{\text{bark}} = 0.5 \left(C_{ij}\right)_{\text{trunk}}$ & 0 &	0.108 & 5.713 \\  \hline \hline
      \multicolumn{4} {|c|}{(b) Variation in bark growth strains} \\ \hline \hline
    (iii) $k_{3\theta}=k_{3r}=-0.0002$, $k_{3z}=-0.0009$ &	0  & 0.213  &	10.999 \\  \hline
 (iv) $k_{3\theta}=k_{3r}=-0.0002$, $k_{3z}=0.0009$  & 0 &	-0.284 &	-8.637 \\  \hline 
  (v) $k_{3\theta}=k_{3r}=-0.0009$, $k_{3z}=-0.0009$  & 0 &	0.633 &	11.192  \\  \hline
(vi) $k_{3\theta}=k_{3r}=0.0009$, $k_{3z}=-0.0009$  & 0 &	-0.447 &	10.697  \\  \hline
(vii) $k_{3\theta}=k_{3r}=-0.009$, $k_{3z}=0.0002$  & 0 &	0.329 &	-0.808  \\  \hline
  \end{tabular}
  \caption{Variation of the stress state in the bark with varying (a) stiffness and (b) bark growth strains.}
  \label{Table-1}
\end{table}

\section{Nutrient concentration during cutaneous wound healing}
\label{wh}
\begin{figure}[t!]
\centering
 \subfloat[]{\includegraphics[width=0.44\textwidth]{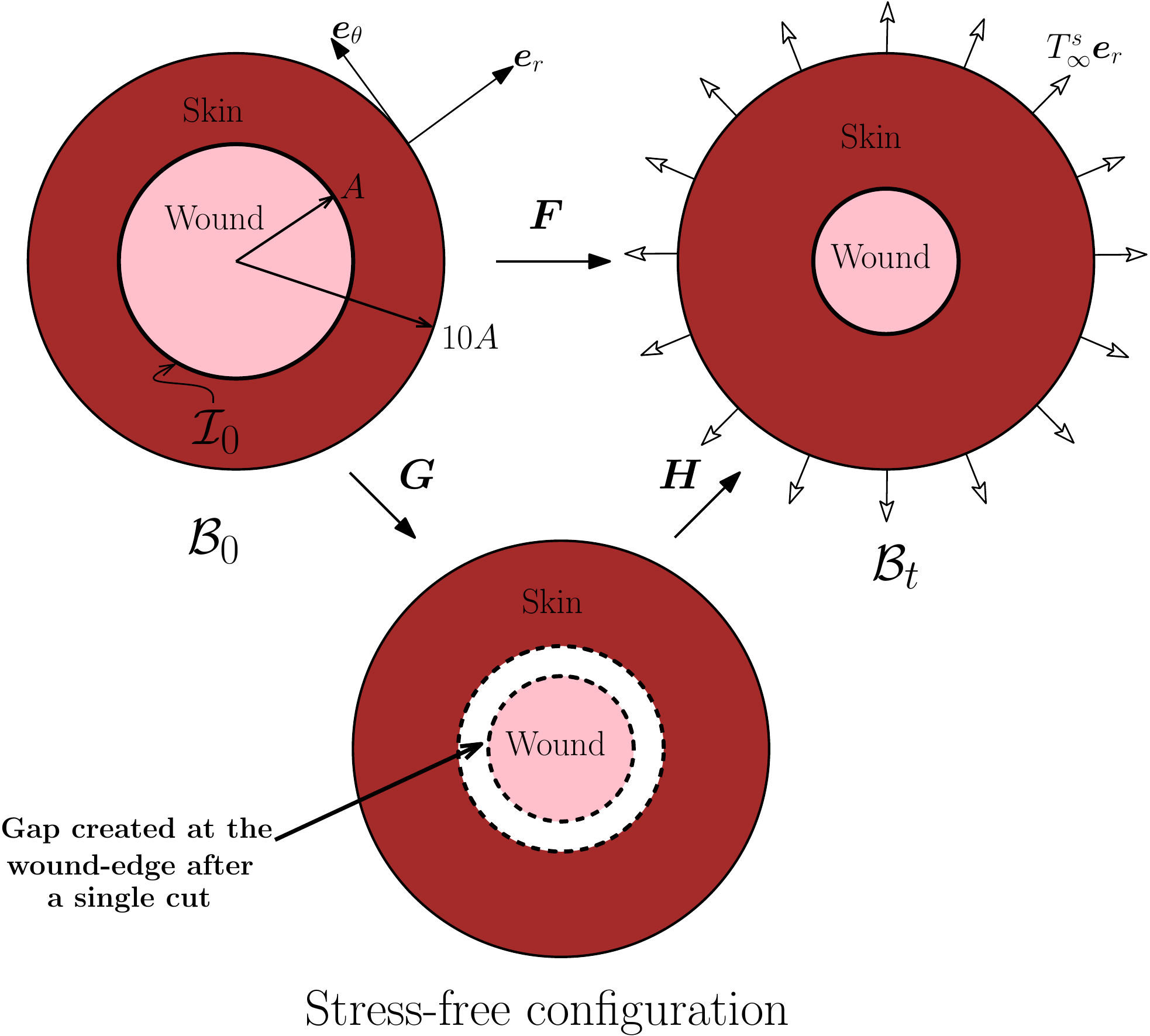}}
 \hspace{-1mm}
 \subfloat[]{\includegraphics[width=0.55\textwidth]{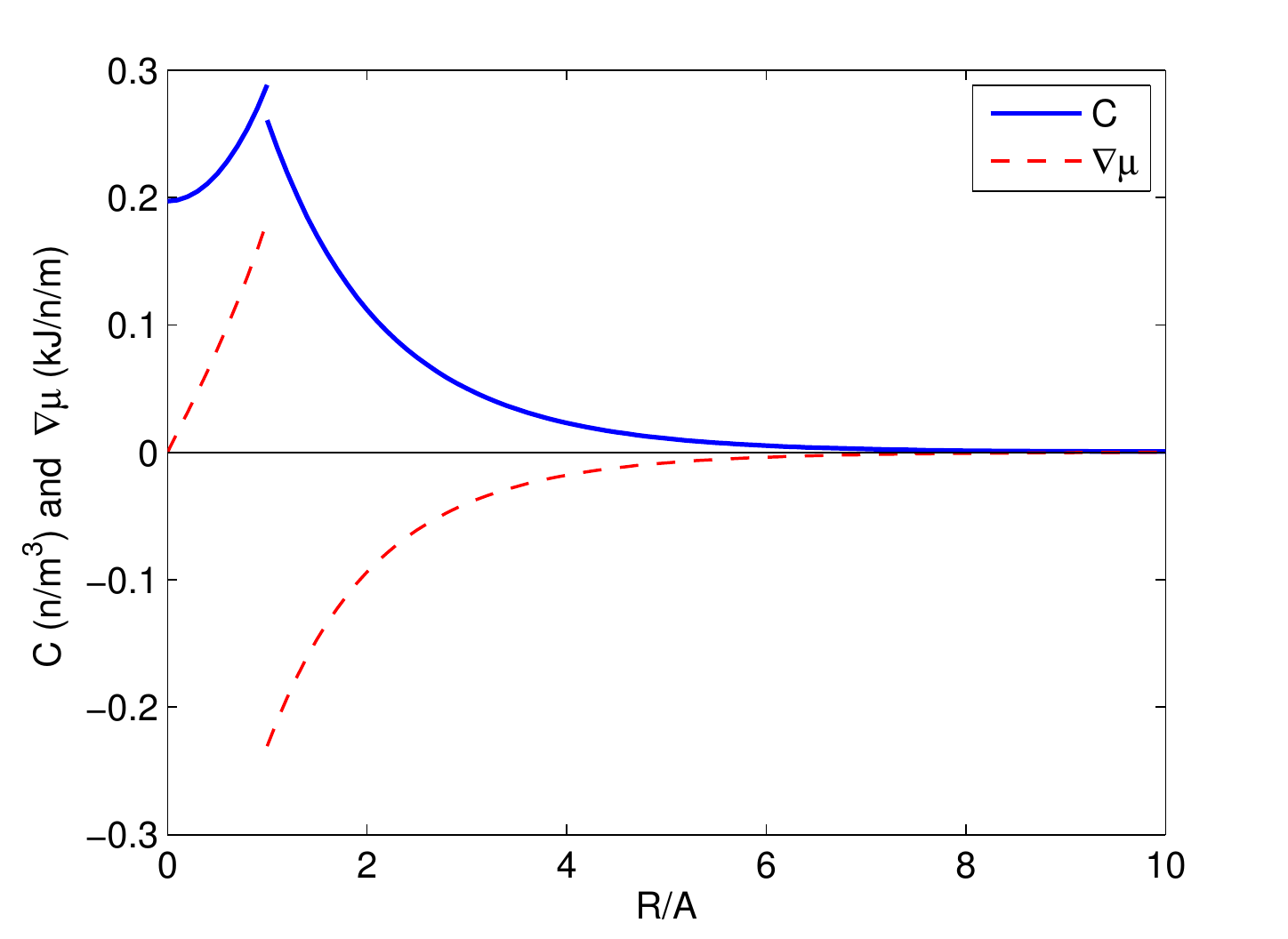}}
\caption{(a) The wound-skin configurations. (b) The variation of nutrient concentration and flux in the wound-skin domain; n$=10^{15}$ cells.}
\label{woundplots}
\end{figure}

We have recently proposed a biomechanical growth model for the proliferation stage of cutaneous wound healing while emphasizing the residual stress generation and the emergence of wrinkling and cavitation instabilities \cite{swain15, swain16}. In this section, we will use an unwrinkled stress solution from our previous work to solve the nutrient balance equations and obtain the steady state nutrient concentration field in the skin-wound bulk region and at the wound edge. A mass source is considered at the incoherent interface $\mathcal{I}_0$ between the circular wound domain and the infinite annular skin domain, see Figure \ref{woundplots}(a), so as to compensate for the density difference between the wound and skin. Both the domains are modeled as isotropic hyperelastic Varga membranes. The stress free configuration is obtained by making a single cut along the wound edge and relaxing the existing far field tension in the skin. 

The problem is considered to be axisymmetric, yielding a deformation gradient of the form $\boldsymbol{F}=r'(R) \boldsymbol{e}_r \otimes \boldsymbol{e}_r+(r(R)/R) \boldsymbol{e}_\theta \otimes \boldsymbol{e}_\theta + (h(R)/H) \boldsymbol{e}_z \otimes \boldsymbol{e}_z$, where $r$ and $R$ are, respectively, the deformed and the reference radial coordinate, $H$ is the uniform thickness of the reference membrane, and $h$ is the thickness of the deformed membrane. The growth deformation is taken to be piecewise uniform, $\boldsymbol{G}=k_a(\boldsymbol{e}_r \otimes \boldsymbol{e}_r+ \boldsymbol{e}_\theta \otimes \boldsymbol{e}_\theta) + \boldsymbol{e}_z \otimes \boldsymbol{e}_z$, where $k_a=k_1$ for the wound region and $k_a=k_2$ for the skin region. The wound-skin interface is incoherent if $k_1 \neq k_2$; in fact, we require $k_2 > k_1$ for a healing wound \cite{swain15}. The evolution of the parameters $k_a$ will be driven both by biochemistry and elastic stresses as is evident from the kinetic laws in \eqref{eq:loc_dissp_8}$_1$. Rather than solving the fully coupled system of equations for deformation and concentration, we will restrict ourselves to obtain a steady state decoupled solution for the concentration, via Equations  \eqref{eq:conc2}, using a known unwrinkled deformation solution from \citet[$\S 2.4$]{swain15}. Accordingly, we will consider $\dot{C}=0$, $\mathring{\mathbb{C}}=0$, $U=0$, and assume the deformation $r$ and the stress fields to be known.

There is no mass addition except at the wound edge. Therefore, $\tr{\dot{\boldsymbol{G}} \boldsymbol{G}^{-1}}=0$ away from the interface. In accordance with the kinetic relation \eqref{eq:loc_dissp_8}$_1$, the tensor $ \boldsymbol{E}_0$, which controls the source of nutrient concentration, should satisfy $\tr {(\widetilde{\boldsymbol{E}} + \mu \boldsymbol{E}_0) } =0$. We take it to be such that $\mu \boldsymbol{E}_0=-(\tr \widetilde{\boldsymbol{E}}) \boldsymbol{e}_r \otimes  \boldsymbol{e}_r$. The steady state form of the nutrient balance law \eqref{eq:conc2}$_1$, with substitutions from \eqref{eq:loc_dissp_8} and choosing $\boldsymbol{K}_0 = K_0 \boldsymbol{1}$, then reduces to $\mu K_0 \triangle \mu = g(C) (\tr{\widetilde{\boldsymbol{E}}}) (\widetilde{{E}}_{\theta \theta} + \widetilde{{E}}_{zz})$, where $\triangle$ is the Laplacian operator. The free energy densities, per unit volume of the stress-free configuration, of the wound and the skin membrane are additively composed of a Varga strain energy \cite{swain15} and a quadratic dependence on the concentration field. The chemical potential in the bulk is therefore given by $\mu (R)= J_G \alpha_a C(R)$, where the material constant $\alpha_a$ is equal to $\alpha_1$ in the wound and to $\alpha_2$ in the skin. The nutrient balance equation consequently takes the form
\begin{equation}
\label{wheq1}
 {K}_0 \alpha_a^2 k_a^4 C \left( C''(R) + \frac{1}{R} C'(R) \right) = g(C)  (\tr{\widetilde{\boldsymbol{E}}}) (\widetilde{{E}}_{\theta \theta} + \widetilde{{E}}_{zz}) ~\text{in}~\mathcal{B}_0/\mathcal{I}_0,
\end{equation}
both within the wound and the skin domain ($a=1$ and $a=2$, respectively). Additionally, chemical equilibrium at the interface requires $\mu^+ = \mu^-$, where the superscripts denote the limiting values of the field at the wound-skin interface with unit normal to the surface pointing into the skin domain. As a result, 
\begin{equation}
\label{wheq2}
k_1^2 \alpha_1 C^- = k_2^2 \alpha_2 C^+ ~\text{on}~\mathcal{I}_0.
\end{equation}

We neglect elastic contributions in the wound-skin interfacial free energy density, per unit area of the stress-free configuration, taking it to be of the form $\widetilde{\Psi}_S = (\beta/2) \mathbb{C}^2$, where $\beta$ is a material constant. The kinetic relations \eqref{eq:loc_dissp_15}$_1$ and \eqref{eq:loc_dissp_15}$_2$ hence reduce to $ \mathring{\mathbb{G}}^\gamma(\mathbb{G}^\gamma)^{-1}= -h_2(\mathbb{C}) \left( j^\gamma \widetilde{\Psi}_S \mathbbm{1}+ \mu {\mathbb{E}}_0^\gamma \right) $ and $ \mathring{\mathbb{G}}^\eta(\mathbb{G}^\eta)^{-1} = -h_1(\mathbb{C}) \left(  \mu {\mathbb{E}}_0^\eta \right)$, respectively.
Substituting these into the nutrient balance equation \eqref{eq:conc2}$_2$, with $\mathring{\mathbb{C}}=0$, $U=0$ (steady state), and $\mathbb{M}=\boldsymbol{0}$ (no intrinsic flux), we obtain ${K}_0 \llbracket \Grad \mu \rrbracket \cdot \boldsymbol{e}_r =  - \sum_{\alpha \in \{\gamma, \eta\}} {\mathbb{E}}_0^\alpha \cdot \mathring{\mathbb{G}}^\alpha (\mathbb{G}^\alpha)^{-1}$. The chemical potential for the interface is equal to the limiting values of the bulk potential, hence $k_2 \beta \mathbb{C} = \alpha_1 k_1^2  C^-$.
Additionally, if we assume $\mathbb{E}_0^{\alpha} =-\mathbbm{1}$ n/m$^2$ (n $=10^{15}$ cells) and $h_1=h_2$, then the interfacial nutrient balance yields
\begin{equation} 
\label{wheq3}
{K}_0 \left (\alpha_2 k_2^2 C'(R)^+ -  \alpha_1 k_1^2 C'(R)^-\right)    =  \tilde{h} (C^-) \left(- \frac{\alpha_1^2 k_1^4}{k_2 \beta} (C^-)^2 + 4  \alpha_1 k_1^2 C^-\right) ~\text{on}~\mathcal{I}_0,
\end{equation}
where $\tilde{h} (C^-)=h_1(\mathbb{C})$. 

The complete boundary-value problem requires solving the nonlinear second-order differential equation \eqref{wheq1} in the wound and the skin domain. The two boundary conditions require the concentration gradient $C'(R)$ to vanish both at the wound center ($R=0$) and far away from the wound edge in the skin, say at $R=10A$, where $A$ is the reference wound radius. In addition, we have two conditions at the interface given by the chemical equilibrium condition \eqref{wheq2} and the interfacial nutrient balance \eqref{wheq3}. For computational purposes, we consider ${K}_{0}=1$ n$^2$/kJ-m-s, $\alpha_1=\alpha_2=1$ kJm$^3$/n$^2$, $g(C)=C^2$, $ \tilde{h}({C}^-)=( k_1^2 C^- / k_2 )^2$, and $\beta=0.01$ kJm$^2$/n$^2$. The deformation and the stress solution is obtained following \citet[$\S 2.4$]{swain15}, where the pre-stress $T^s_\infty$ at the outer boundary of the skin region is taken as $0.477634$ kPa, skin stiffness as $8$ kPa, wound stiffness $5.6$ kPa, $k_1=0.9602$, $k_2=1.01$, and the healing constant as $0.99$.
The problem is solved numerically using an iterative procedure with results summarized in Figure \ref{woundplots}(b). That the nutrient concentration is larger on the wound side of the interface is indicative of the active cellular activities therein. The chemical potential gradients are sharp near the wound edge implying that these results are a snapshot of an initial instant in the process of wound healing.

\section{Conclusion}
\label{conc}
We have developed a framework for studying biological growth in bodies with incoherent interfaces. The main contributions of our work include incorporating the incoherency into interfacial growth models and to couple the nutrient concentration evolution with the growth evolution, both in the bulk and at the interface. The incoherency of the interface leads to internal stress in the body and allows for growth distortions to evolve across the interface without necessarily being compatible. The nutrient concentration and the growth evolutions, in addition to the momentum balances and the constitutive relations, form the complete set of governing equations to be solved for displacement, stress, and concentration. This was illustrated through simplified models of tree growth and wound healing, where valuable insights were obtained into the biomechanical and biochemical aspects of these problems. Suitable numerical strategies would have to be devised to deal with more sophisticated problems. Additionally, the proposed model would have to find other applications in surface and interfacial growth, for instance in problems of nail and bone growth. Our interface growth model can also be explored to provide meaningful boundary conditions for a higher-gradient bulk growth theory. 

\medskip

\noindent {\bf Data accessibility}. The article has no additional data. 

\noindent{\bf Acknowledgement}. We thank Prof. Sovan Das for his valuable comments.

\noindent {\bf Funding statement}. Neither author received funding to carry out this research. 

\noindent {\bf Authors' contributions}. A.G. planned the research. D.S. conducted the research, and worked out all the derivations and examples. Both A.G. and D.S. analyzed the results and contributed equally to the writing.

\noindent {\bf Competing interests}. The authors have no competing interests.

\citestyle{numbers}
\bibliographystyle{plainnat}
\bibliography{ag_revised}

\begin{thebibliography}{34}
\providecommand{\natexlab}[1]{#1}
\providecommand{\url}[1]{\texttt{#1}}
\expandafter\ifx\csname urlstyle\endcsname\relax
  \providecommand{\doi}[1]{doi: #1}\else
  \providecommand{\doi}{doi: \begingroup \urlstyle{rm}\Url}\fi

\bibitem[Ambrosi and Guillou(2007)]{ambrosi07}
D.~Ambrosi and A.~Guillou.
\newblock Growth and dissipation in biological tissues.
\newblock \emph{Contin. Mech. Thermodyn.}, 19:\penalty0 245--251, 2007.

\bibitem[Ambrosi and Mollica(2002)]{ambrosi02}
D.~Ambrosi and F.~Mollica.
\newblock On the mechanics of a growing tumor.
\newblock \emph{Internat. J. Engrg. Sci.}, 40:\penalty0 1297--1316, 2002.

\bibitem[Archer(1986)]{archer86}
R.~R. Archer.
\newblock \emph{Growth Stresses and Strains in Trees}.
\newblock Springer-Verlag Berlin Heidelberg, 1986.

\bibitem[Basak and Gupta(2015)]{basak15b}
A.~Basak and A.~Gupta.
\newblock A three-dimensional study of coupled grain boundary motion with
  junctions.
\newblock \emph{Proc. R. Soc. Lond. Ser. A Math. Phys. Eng. Sci.},
  471:\penalty0 20150127, 2015.

\bibitem[Basak and Gupta(2016)]{basak15a}
A.~Basak and A.~Gupta.
\newblock Plasticity in multi-phase solids with incoherent interfaces and
  junctions.
\newblock \emph{Contin. Mech. Thermodyn.}, 28:\penalty0 423--442, 2016.

\bibitem[Buskohl et~al.(2014)Buskohl, Butcher, and Jenkins]{buskohl14}
P.~R. Buskohl, J.~T. Butcher, and J.~T. Jenkins.
\newblock The influence of external free energy and homeostasis on growth and
  shape change.
\newblock \emph{J. Mech. Phys. Solids}, 64:\penalty0 338 -- 350, 2014.

\bibitem[Ciarletta et~al.(2013)Ciarletta, Preziosi, and Maugin]{ciarletta12}
P.~Ciarletta, L.~Preziosi, and G.~A. Maugin.
\newblock Mechanobiology of interfacial growth.
\newblock \emph{J. Mech. Phys. Solids}, 61:\penalty0 852--872, 2013.

\bibitem[DiCarlo(2005)]{dicarlo05}
A.~DiCarlo.
\newblock Surface and bulk growth unified.
\newblock In P.~Steinmann and G.~A. Maugin, editors, \emph{Mechanics of
  Material Forces}, pages 53--64. Springer US, 2005.

\bibitem[Epstein and Maugin(2000)]{epstein2000}
M.~Epstein and G.~A. Maugin.
\newblock Theromechanics of volumetric growth in uniform bodies.
\newblock \emph{Internat. J. Plast.}, 16:\penalty0 951--978, 2000.

\bibitem[Fournier et~al.(1990)Fournier, Bordonne, Guitad, and
  Okuyama]{fournier90}
M.~Fournier, P.~A. Bordonne, D.~Guitad, and T.~Okuyama.
\newblock Growth stress patterns in tree stems.
\newblock \emph{Wood Sci. Technol.}, 24:\penalty0 131--142, 1990.

\bibitem[Ganghoffer(2011)]{ganghoffer11}
J.~F. Ganghoffer.
\newblock Mechanics and thermodynamics of surface growth viewed as moving
  discontinuities.
\newblock \emph{Mech. Res. Commun.}, 38:\penalty0 372 -- 377, 2011.

\bibitem[Giuseppe and Truskinovsky(2017)]{truskinovsky17}
Z.~Giuseppe and L.~Truskinovsky.
\newblock Printing non-euclidean solids.
\newblock \emph{Phys. Rev. Lett.}, 119:\penalty0 048001, 2017.

\bibitem[Goriely(2017)]{Goriely17}
A.~Goriely.
\newblock \emph{The mathematics and mechanics of biological growth}.
\newblock Springer-Verlag New York, 2017.

\bibitem[Gupta and Steigmann(2012)]{Gupta12}
A.~Gupta and D.~J. Steigmann.
\newblock Plastic flow in solids with interfaces.
\newblock \emph{Math. Methods Appl. Sci.}, 35:\penalty0 1799--1824, 2012.

\bibitem[Gupta et~al.(2007)Gupta, Steigmann, and St{\"o}lken]{Guptaetal07}
A.~Gupta, D.~J. Steigmann, and J.~St{\"o}lken.
\newblock On the evolution of plasticity and incompatibility.
\newblock \emph{Math. Mech. Solids}, 12:\penalty0 583--610, 2007.

\bibitem[Holland et~al.(2013)Holland, Kosmata, Goriely, and Kuhl]{kuhl13b}
M.~A. Holland, T.~Kosmata, A.~Goriely, and E.~Kuhl.
\newblock On the mechanics of thin films and growing surfaces.
\newblock \emph{Math. Mech. Solids}, 18:\penalty0 561--575, 2013.

\bibitem[Jones and Chapman(2012)]{jones12}
G.~W. Jones and S.~J. Chapman.
\newblock Modeling growth in biological materials.
\newblock \emph{SIAM Review}, 54:\penalty0 52--118, 2012.

\bibitem[Mattheck and Kubler(1997)]{mattheck97}
C.~Mattheck and H.~Kubler.
\newblock \emph{Wood -The Internal Optimization of Trees}.
\newblock Springer-Verlag Berlin Heidelberg, 1997.

\bibitem[Moulton and Goriely(2011)]{moulton11b}
D.~E. Moulton and A.~Goriely.
\newblock Circumferential buckling instability of a growing cylindrical tube.
\newblock \emph{J. Mech. Phys. Solids}, 59:\penalty0 525--537, 2011.

\bibitem[Oller et~al.(2010)Oller, Bellomo, Armero, and Nallim]{oller10}
S.~Oller, F.~J. Bellomo, F.~Armero, and L.~G. Nallim.
\newblock A stress driven growth model for soft tissue considering biological
  availability.
\newblock \emph{IOP Conf. Series: Materials Science and Engineering},
  10\penalty0 (012121):\penalty0 1--10, 2010.

\bibitem[Pallardy(2008)]{pallardy08}
S.~G. Pallardy.
\newblock \emph{Physiology of Woody Plants}.
\newblock Elsevier, third edition, 2008.

\bibitem[Papastavrou et~al.(2013)Papastavrou, Steinmann, and Kuhl]{kuhl13a}
A.~Papastavrou, P.~Steinmann, and E.~Kuhl.
\newblock On the mechanics of continua with boundary energies and growing
  surfaces.
\newblock \emph{J. Mech. Phys. Solids}, 61:\penalty0 1446--1463, 2013.

\bibitem[Plomion et~al.(2001)Plomion, Leprovost, and Stokes]{plomion01}
C.~Plomion, G.~Leprovost, and A.~Stokes.
\newblock Wood formation in trees.
\newblock \emph{Plant Physiol.}, 127:\penalty0 1513--1523, 2001.

\bibitem[Post et~al.(1980)Post, Atherton, Vendhan, and Archer]{Post80}
I.~L. Post, J.~C. Atherton, C.~P. Vendhan, and R.~R. Archer.
\newblock An extension of {J}acobs' method for measuring residual growth
  strains in logs.
\newblock \emph{Wood Sci. Technol.}, 14:\penalty0 289--296, 1980.

\bibitem[Rodriguez et~al.(1994)Rodriguez, Hoger, and Mcculloch]{Rodriguez1994}
E.~K. Rodriguez, A.~Hoger, and A.~D. Mcculloch.
\newblock Stress-dependent finite growth in soft elastic tissues.
\newblock \emph{J. Biomech.}, 21:\penalty0 455--467, 1994.

\bibitem[Roychowdhury and Gupta(2017)]{Ayan16}
A.~Roychowdhury and A.~Gupta.
\newblock Non-metric connection and metric anomalies in materially uniform
  elastic solids.
\newblock \emph{J. Elas.}, 126:\penalty0 1--26, 2017.

\bibitem[Skalak et~al.(1997)Skalak, Farrow, and Hoger]{Skalak1997}
R.~Skalak, D.~A. Farrow, and A.~Hoger.
\newblock Kinematics of surface growth.
\newblock \emph{J. Math. Biol.}, 35:\penalty0 869--907, 1997.

\bibitem[Sugiyama et~al.(1993)Sugiyama, Okuyama, Yamamoto, and
  Yoshida]{sugiyama93}
K.~Sugiyama, T.~Okuyama, H.~Yamamoto, and M.~Yoshida.
\newblock Generation process of growth stresses in cell walls: Relation between
  longitudinal released strain and chemical composition.
\newblock \emph{Wood Sci. Technol.}, 27:\penalty0 257--262, 1993.

\bibitem[Swain and Gupta(2015)]{swain15}
D.~Swain and A.~Gupta.
\newblock Interfacial growth during closure of a cutaneous wound: Stress
  generation and wrinkle formation.
\newblock \emph{Soft Matter}, 11:\penalty0 6499--6508, 2015.

\bibitem[Swain and Gupta(2016)]{swain16}
D.~Swain and A.~Gupta.
\newblock Mechanics of cutaneous wound rupture.
\newblock \emph{J. Biomech.}, 49:\penalty0 3722--3730, 2016.

\bibitem[Taber(1995)]{taber95}
L.~A. Taber.
\newblock Biomechanics of growth, remodeling, and morphogenesis.
\newblock \emph{Ann. Biomed. Eng.}, 48:\penalty0 487--545, 1995.

\bibitem[Wilkins(1986)]{wilkins86}
A.~P. Wilkins.
\newblock Nature and origin of growth stresses in trees.
\newblock \emph{Aust. For.}, 49:\penalty0 56--62, 1986.

\bibitem[Wolff and B\"{o}hn(2017)]{wolff15}
M.~Wolff and M.~B\"{o}hn.
\newblock Continuous bodies with thermodynamically active singular sharp
  interfaces.
\newblock \emph{Math. Mech. Solids}, 22:\penalty0 434--476, 2017.

\bibitem[Wu and Amar(2015)]{WuAmar15}
M.~Wu and M.~B. Amar.
\newblock Growth and remodelling for profound circular wounds in skin.
\newblock \emph{Biomech. Model. Mechanobiol}, 14:\penalty0 357--370, 2015.

\end{thebibliography}

\end{document}